\newif\ifonecol

\onecoltrue    % For one column document
\ifonecol
\documentclass[11pt,draftcls,onecolumn,peerreview]{IEEEtran}
\else
\documentclass[journal]{IEEEtran}
\fi

\usepackage{amsfonts}
\usepackage{amsmath}
\usepackage{amssymb}
\usepackage[final]{graphicx}
\usepackage{xspace}
\usepackage{intmacros}
\usepackage{subfigure}
\newcommand{\Nu}{m}  % Number of Unknowns = (in SCA:) Number of Sources = (in Atomic Decomp:) Number of atoms
\newcommand{\Ne}{n}  % Number of Equations = (in SCA:) Number of Mixtures = (in Atomic Decomp:) Length of the signal
\newcommand{\xb}{{\bf x}}
\newcommand{\ab}{{\bf a}}

\newcommand{\nb}{{\bf n}}
\newcommand{\tnb}{{\bf \tilde n}}
\newcommand{\tsb}{{\bf \tilde s}}
\newcommand{\hsb}{{\bf \hat s}}
\newcommand{\hAb}{{\bf \hat{\Ab}}}
\newcommand{\bsb}{{\bf \bar{s}}}
\newcommand{\sbb}{{\bf s}}
\newcommand{\Ab}{{\bf A}}

\newcommand{\Sb}{{\bf S}}
\newcommand{\Xb}{{\bf X}}
\newcommand{\Nb}{{\bf N}}
\newcommand{\Ib}{{\bf I}}

\newcommand{\vb}{{\bf v}}

\newcommand{\sigJ}{{\sigma_{\!J}}}
\newcommand{\lambdab}{\boldsymbol{\lambda}}

\newcommand{\deltab}{\boldsymbol{\delta}}

\newcommand{\oureq}{\Ab \sbb = \xb}
\newcommand{\Sm}{{\mathcal{S}}}

\newcommand\eg{e.g.\xspace}
\newcommand\ie{i.e.,\xspace}
\newcommand{\oursystem}{\Ab \sbb = \xb}
\newcommand{\Rr}{\mathbb{R}}

\newcommand{\Fs}{F_\sigma}
\newcommand{\fs}{f_\sigma}

\newcommand{\norm}[2]{\|#1\|_{#2}}
\newcommand{\Lzero}{$\ell^0$}
\newcommand{\Lone}{$\ell^1$}
\newcommand{\Ltwo}{$\ell^2$}

\newtheorem{theorem}{Theorem}
\newtheorem{lemma}{Lemma}
\newtheorem{coro}{Corollary}

\newcommand{\inst}[1]{\unskip${^{#1}}$}

\title{A fast approach for overcomplete sparse decomposition based on smoothed \Lzero\ norm}
 \author{Hosein Mohimani \inst{1}, Massoud
  Babaie-Zadeh\inst{1}*~\IEEEmembership{Member}  and Christian Jutten \inst{2}~\IEEEmembership{Fellow}
  \thanks{$^1$Electrical engineering department, Sharif university of
    technology, Tehran, Iran.}
  \thanks{$^2$GIPSA-Lab, Department of Images and Signals,
    Institut National Polytechnique de Grenoble (INPG), France.}
  \thanks{This work has been partially supported by Iran National Science Foundation (INSF) under contract number 86/994, and also by center for International Research and Collabration (ISMO) and French embassy in Tehran in the framework of a GundiShapour program.}
  \thanks{Author's email addresses are: {\tt gh1985im@yahoo.com}, {\tt mbzadeh@yahoo.com} and {\tt Christian.Jutten@inpg.fr}}
  \ifonecol
  \thanks{Corresponding author: Massoud Babaie-Zadeh, email: {\tt mbzadeh@yahoo.com}, Tel: +98 21 66 16 59 25, Fax: +98 21 66 02 32 61.}
  \fi
  }

\begin{document}
\maketitle

\begin{abstract}
In this paper, a fast algorithm for overcomplete sparse decomposition, called SL0, is
proposed. The algorithm is essentially a method for obtaining
sparse solutions of underdetermined systems of linear equations,
and its applications include underdetermined Sparse Component
Analysis (SCA), atomic decomposition on overcomplete dictionaries,
compressed sensing, and decoding real field codes. Contrary to
previous methods, which usually solve this problem by minimizing the
\Lone\ norm using Linear Programming (LP) techniques, our
algorithm tries to directly minimize the \Lzero\ norm. It is
experimentally shown that the proposed algorithm is about two
to three orders of magnitude faster than the state-of-the-art
interior-point LP solvers, while providing the same (or better)
accuracy.
\end{abstract}

%%%%%%%%%%%%%%%%%%%%%%%%%%%%%%%%%%%%%%%%%%%%%%%%%%%%%%%%%%%%%%%%%
\begin{keywords}
Sparse decomposition, compressed sensing, Sparse Component Analysis (SCA), atomic
decomposition, overcomplete signal representation, sparse source
separation, Blind Source Separation (BSS).
\end{keywords}

%%%%%%%%%%%%%%%%%%%%%%%%%%%%%%%%%%%%%%%%%%%%%%%%%%%%%%%%%%%%%%%%%
\ifonecol

\begin{center} \bfseries EDICS Category: MAL-SSEP, SPC-CODC, DSP-FAST \end{center}

\IEEEpeerreviewmaketitle

\fi

%%%%%%%%%%%%%%%%%%%%%%%%%%%%%%%%%%%%%%%%%%%%%%%%%%%%%%%%%%%%%%%%%%%
%%%%%%%%%%%                 New Section                 %%%%%%%%%%%
%%%%%%%%%%%%%%%%%%%%%%%%%%%%%%%%%%%%%%%%%%%%%%%%%%%%%%%%%%%%%%%%%%%

\label{sec: int}

\section{Introduction}
\PARstart{F}{inding sparse} solutions of Under-determined Systems
of Linear Equations (USLE) is of significant importance in signal
processing and statistics. It is used, for example, in
underdetermined Sparse Component Analysis (SCA) and source
separation~\cite{GribL06, BofiZ01, GeorTC04, LiCA03}, atomic decomposition on
overcomplete dictionaries~\cite{ChenDS99,DonoET06}, compressed
sensing~\cite{Dono06,Bara07}, decoding real field
codes~\cite{CandT05}, image deconvolution~\cite{FiguN03,FiguN05},
image denoising~\cite{Elad06}, electromagnetic imaging and
Direction of Arrival (DOA) finding~\cite{GoroR97}. Despite recent
theoretical
developments~\cite{DonoE03,Dono04,DaubDD04,CandRT06}, the
computational cost of the methods has remained as the main
restriction, especially for large systems (large number of
unknowns/equations). In this article, a new approach is proposed
which provides a considerable reduction in complexity. To
introduce the problem in more details,  we will use the context
of Sparse Component Analysis (SCA). The discussions, however, may
be easily followed in other contexts and applications.

SCA can be viewed as a method to achieve separation of sparse
sources. Suppose that $m$ source signals are recorded by a set of $n$
sensors, each of which records a combination of all sources. In linear
instantaneous (noiseless) model, it is assumed
that $\xb(t) = \Ab \sbb(t)$ in which $\xb(t)$ and $\sbb(t)$ are
the $\Ne\times1$ and $\Nu\times1$ vectors of source and recorded signals, respectively,
and $\Ab$ is the $\Ne \times \Nu$ (unknown) mixing matrix. The goal of
Blind Source Separation (BSS)~\cite{HyvaKO01,CichA02}
is then to find $\sbb(t)$ only by observing $\xb(t)$. The general
BSS problem is impossible for the case $\Nu>\Ne$. However, if the
sources are sparse (\ie not a totally blind situation), then the
problem can be solved in two steps \cite{GribL06,BofiZ01}: first
estimating the mixing matrix, and then estimating the sources
assuming $\Ab$ being known. For sparse sources, the first step 
-- which can become very tricky for large $m$ -- may be accomplished
by means of clustering
\cite{GribL06,BofiZ01,MovaMBJ06,WashC06}. The second step requires
that for each sample ($t_0$) the sparse solution of the USLE
$\xb(t_0) = \Ab \sbb(t_0)$ be found~\cite{GribL06,BofiZ01,LiCA04,ZibuP01}.
Note also that the sparsity of the sources is not necessarily in the time domain: if
$T\{.\}$ is a linear `sparsifying' transformation, then $T\{\xb\}=\Ab \, T \{\sbb \}$. Due to linearity of $T$, both the linearity of the mixing and the statistical independence properties of sources are preserved in the transformed domain. Hence, SCA may be applied in the transformed domain.

In the atomic decomposition viewpoint \cite{ChenDS99},
the signal vector $\xb=[x(1),\dots,x(\Ne)]^T$ is composed of the samples of a `single' signal $x(t)$, and the objective is to represent it as a linear combination of $\Nu$, $\Ne \times
1$ signal vectors $\{\ab_{i}\}_{i=1}^{\Nu}$. After
\cite{MallZ93}, the vectors $\ab_{i}$, $1 \leq i \leq \Nu$ are
called \emph{atoms} and they collectively form a
\emph{dictionary} over which the signal is to be decomposed. We
may write $\xb = \sum_{i=1}^{\Nu} s_{i} \ab_{i} = \Ab
\sbb$, where $\Ab \triangleq [\ab_{1},\dots,\ab_{\Nu}]$ is
the $\Ne \times \Nu$ dictionary (matrix) and $\sbb \triangleq
(s_1,\dots,s_\Nu)^T$ is the $\Nu \times 1$ vector of
coefficients. A dictionary with $\Nu > \Ne$ is called
\emph{overcomplete}. Although, $\Nu = \Ne$ (\eg Discrete Fourier Transform)
is sufficient to obtain such a decomposition, using
overcomplete dictionaries has a lot of advantages in many diverse
applications (refer for example to \cite{DonoET06} and the
references in it). In all these applications, we would like to
use as small as possible number of atoms to represent the signal.
Again, we have the problem of finding sparse solutions of the
USLE $\Ab \sbb= \xb$.

To obtain the sparsest solution of $\oursystem$, we may search
for a solution with minimal \Lzero\ norm, \ie minimum
number of nonzero components. It is usually stated in the
literature \cite{DonoET06,CandT05,LiCA03} that searching the
minimum \Lzero\ norm is an intractable problem as the
dimension increases (because it requires a combinatorial search),
and it is too sensitive to noise (because any small amount of
noise completely changes the \Lzero\ norm of a vector).
Consequently, researchers consider other approaches. One of
the most successful approaches is Basis Pursuit (BP)
\cite{ChenDS99,Dono04,LiCA03, CandR05} which finds the minimum
\Lone\ norm (that is, the solution of $\oursystem$ for which
$\sum_i{|s_i|}$ is minimized). Such a solution can be easily
found by Linear Programming (LP) methods. The idea of Basis
Pursuit is based on the observation that for large systems of
equations, the minimum \Lone\ norm solution is also the minimum
\Lzero\ norm solution \cite{DonoE03,Dono04,ChenDS99}. By
using fast LP algorithms, specifically interior-point LP
solvers, large-scale problems with thousands of sources and
mixtures become tractable. However, it is still very slow, and in
the recent years several authors have proposed improvements for BP,
to speed up the algorithm and to handle the noisy
case~\cite{DaubDD04,DonoET06,FiguN03,FiguN05}. Another family of
algorithms is Iterative Re-weighted Least Squares (IRLS), with
FOCUSS~\cite{GoroR97} as an important member. These are faster
than BP, but their estimation quality is worse, especially if
the number of non-zero elements of the sparsest solution is not
very small. Another approach is Matching Pursuit (MP)
\cite{MallZ93,KrstG06,GribL06} which is very fast, but is a
greedy algorithm and does not provide good estimation of the
sources. The approach presented in~\cite{AminBJ06} is also very
fast, but adjusting its parameters is not easy.

Contrary to previous approaches, the method we present in this
paper is based on direct minimization of the \Lzero\ norm. We
will see that our method performs typically {\em two to three orders of
magnitude faster than BP\/} (based on interior-point LP solvers),
while resulting in the same or better accuracy. We have
already briefly reported the basics of this approach in~\cite{MohiBJ07}
and its complex version in~\cite{MohiBJ08}. However, in this paper,
we are going to present a highly more complete description of this approach
and consider, mathematically and/or experimentally,
its convergence properties and the effects of its parameters.

The paper is organized as follows. The next section introduces
the basic principles of our approach. The final algorithm is then
stated in Section \ref{sec: alg}. In Section~\ref{sec: alg anal},
convergence properties of the algorithm is discussed. Finally,
Section \ref{sec: experiments} provides some experimental results
of our algorithm and its comparison with BP.

%%%%%%%%%%%%%%%%%%%%%%%%%%%%%%%%%%%%%%%%%%%%%%%%%%%%%%%%%%%%%%%%%%%
%%%%%%%%%%%                 New Section                 %%%%%%%%%%%
%%%%%%%%%%%%%%%%%%%%%%%%%%%%%%%%%%%%%%%%%%%%%%%%%%%%%%%%%%%%%%%%%%%
\section{Basic Principles Of Our Approach} \label{sec: Basic Principles}

%%%%%%%%%%%%%%%%%%%%%%%%
%%%  New SubSection  %%%
%%%%%%%%%%%%%%%%%%%%%%%%
\subsection{The Main Idea}

The problems of using \Lzero\ norm (that is, the need for a
combinatorial search for its minimization, and its too high
sensibility to noise) are both due to the fact that the \Lzero\
norm of a vector is a \emph{discontinuous} function of that
vector. Our idea is then to approximate this \emph{discontinuous}
function by a suitable \emph{continuous} one, and minimize it by
means of a minimization algorithm for continuous functions (\eg
steepest descent method). The continuous function which
approximates $\norm{\sbb}{0}$, the \Lzero\ norm of $\sbb$, should
have a parameter (say $\sigma$) which determines the quality of
the approximation.

For example, consider the (one-variable) family of functions:
\begin{equation}\label{eq: Gaussian f}
  f_\sigma(s)\triangleq \exp{(-s^2/2\sigma^2)},
\end{equation}
and note that:
\begin{equation}\label{eq: approximating prop 1}
  \lim_{\sigma \to 0} f_\sigma(s) = \left \{
  \begin{array}{ll}
    1   & ; \mbox{if $s=0$} \\
    0   & ; \mbox{if $s\neq 0$}
  \end{array}
  \right. ,
\end{equation}
or approximately:
\begin{equation}\label{eq: approximating prop 2}
  f_\sigma(s) \approx \left \{
  \begin{array}{ll}
    1   & ; \mbox{if $|s|\ll \sigma$} \\
    0   & ; \mbox{if $|s|\gg \sigma$}
  \end{array}
  \right. .
\end{equation}
Then, by defining:
\begin{equation}\label{eq: Gaussian F}
  F_\sigma(\sbb) = \sum_{i=1}^{\Nu} f_\sigma(s_i),
\end{equation}
it is clear from (\ref{eq: approximating prop 1}) and (\ref{eq:
approximating prop 2}) that $\norm{\sbb}{0} \approx \Nu -
\Fs(\sbb)$ for small values of $\sigma$, and the approximation
tends to equality when $\sigma \to 0$. Consequently, we can find
the minimum \Lzero-norm solution by maximizing $\Fs(\sbb)$
(subject to $\Ab \sbb = \xb$) for a very small value of $\sigma$.
Note that the value of $\sigma$ determines how smooth the
function $\Fs$ is: the larger value of $\sigma$, the smoother
$\Fs$ (but worse approximation to \Lzero-norm); and the smaller
value of $\sigma$, the closer behavior of $\Fs$ to \Lzero-norm.

Note that for small values of $\sigma$, $\Fs$ is highly
non-smooth, and contains a lot of local maxima, and hence its
maximization is not easy. On the other hand, for larger values of
$\sigma$, $\Fs$ is smoother and contains less local maxima, and
its maximization is easier (we will see in the next subsection
that there is no local maxima for large enough $\sigma$'s).
Consequently, our idea is to use a `decreasing' sequence for
$\sigma$: for maximizing $\Fs$ for each value of $\sigma$ (using
\eg gradient algorithms), the initial value of the maximization
algorithm is the maximizer of $\Fs$ for the previous (larger)
value of $\sigma$. If we gradually decrease the value of
$\sigma$, for each value of $\sigma$ the maximization
algorithm starts with an initial solution near to the actual
maximizer of $\Fs$ (this is because $\sigma$ and hence $\Fs$ have
only slightly changed and consequently the maximum of the new $\Fs$
is eventually near to the maximum of the previous $\Fs$),
and hence we eventually escape from getting trapped
into local maxima and reach to the actual maximum for small
values of $\sigma$, which gives the minimum \Lzero-norm
solution\footnote{This technique for optimizing a non-convex
function is usually called Graduated Non-Convexity (GNC)~\cite{BlakZ87}.}.

Note that the basic idea holds not only for Gaussian family of
functions $f_\sigma$ given in (\ref{eq: Gaussian f}), but also for
any family of functions $f_\sigma$ which approximates the Kronecker
delta function, \ie satisfies (\ref{eq: approximating prop 1}) and
(\ref{eq: approximating prop 2}). For example, it also holds for the
family of `triangular' functions:
\begin{equation}\label{eq: trig f}
  f_\sigma(s) = \left \{
  \begin{array}{ll}
    1   & ; \mbox{if $|s| \ge \sigma$} \\
    (\sigma+s)/\sigma   & ; \mbox{if $-\sigma \le s \le 0$}\\
    (\sigma-s)/\sigma   & ; \mbox{if $0 \le s \le \sigma$}
  \end{array}
  \right. ,
\end{equation}
and for the family of `truncated hyperbolic' functions:
\begin{equation}\label{eq: hyperbolic f}
  f_\sigma(s) = \left \{
  \begin{array}{ll}
    1   & ; \mbox{if $|s| \ge \sigma$} \\
    1-(s/\sigma)^2   & ; \mbox{if $|s| \le \sigma$} \\
  \end{array}
  \right. ,
\end{equation}
and also for the family of functions:
\begin{equation} \label{eq: second order mobius f}
  f_\sigma(s) = \sigma^2/(s^2+\sigma^2).
\end{equation}

%%%%%%%%%%%%%%%%%%%%%%%%
%%%  New SubSection  %%%
%%%%%%%%%%%%%%%%%%%%%%%%
\subsection{Initialization} \label{sec: basic init}

Up to now, the behavior of the function $f_{\sigma}$ was
discussed for small values of $\sigma$. It is also interesting to
consider the behavior for very large values of $\sigma$.

More specifically, it can be shown that ``{\em for sufficiently
large values of $\sigma$, the maximizer of $F_{\sigma}(\sbb)$
subject to $\oureq$ is the minimum $\ell^{2}$-norm solution of
$\oureq$, \ie the solution given by the pseudo-inverse of
$\Ab$\/}''. Here, we give only a justification to this property
for the case of Gaussian family of functions introduced in
(\ref{eq: Gaussian f}) by using Lagrange multipliers, and we
leave the formal proof to Section~\ref{sec: exact relation min
norm 2}.

Using the method of Lagrange multipliers, for maximizing
$F_\sigma(\sbb) = \sum_{i=1}^{\Nu} f_\sigma(s_i)= \sum_{i=1}^{\Nu}
\exp{(-s_i^2 /2\sigma^2)}$ subject to $\Ab \sbb = \xb$, we set
the derivative of the Lagrangian ${\cal L}(\sbb,
\lambdab)=F_\sigma(\sbb) - \lambdab^T(\Ab \sbb - \xb)$ with
respect to $\sbb$ and $\lambdab$ equal to zero, which gives the
following Karush-Kuhn-Tucker (KKT) system of $\Nu+\Ne$ nonlinear
equations of $\Nu+\Ne$ unknowns ($\Nu$ components of $\sbb$, and
$\Ne$ components of $\lambdab$):
\begin{equation}\label{eq: theo proof L2}
\left \{
  \begin{array}{ll}
     [s_1 e^{-s_1^2/2\sigma^2},\dots,s_\Nu e^{-s_\Nu^2/2\sigma^2}]^T -
     \Ab^T \lambdab_1 = {\bf 0}\\
     \Ab \sbb - \xb = {\bf
     0}
  \end{array}
  \right.
\end{equation}
where $\lambdab_1 \triangleq - \sigma^2 \lambdab$.

On the other hand, the minimum \Ltwo\ norm solution of $\oureq$ may
be found by minimizing $\frac{1}{2} \sbb^T \sbb$ subject to
$\oureq$. Using Lagrange multipliers, this minimization
results in the system of equations:
\begin{equation}\label{eq: theo proof L3}
\left \{
  \begin{array}{ll}
     [s_1,\dots,s_\Nu]^T -
     \Ab^T \lambdab = {\bf 0}\\
     \Ab \sbb - \xb = {\bf
     0}
  \end{array}
  \right.
\end{equation}
Comparing systems (\ref{eq: theo proof L2}) and (\ref{eq: theo
proof L3}), we see that for $\sigma \to \infty$ (or where $\sigma
\gg \max\{s_1, \dots, s_\Nu\}$), these two systems of equations
are identical, and hence the maximizer of $F_{\sigma}(\sbb)$ is
the minimum $\ell^2$-norm solution of $\oursystem$.

%%%%%%%%%%%%%%%%%%%%%%%%%%%%%%%%%%%%%%%%%%%%%%%%%%%%%%%%%%%%%%%%%%%%
%%%%%%%%%%%%                 New Section                 %%%%%%%%%%%
%%%%%%%%%%%%%%%%%%%%%%%%%%%%%%%%%%%%%%%%%%%%%%%%%%%%%%%%%%%%%%%%%%%%
\section{The Final Algorithm} \label{sec: alg}

The final algorithm, which we call SL0 (Smoothed \Lzero ),
is obtained by applying the main idea of the previous section
on the Gaussian family (\ref{eq: Gaussian f}), and
is given in Fig.~\ref{fig: the alg}.

\vspace{0.5em}

\begin{figure}
  \centering \vrule%
  \ifonecol
    \begin{minipage}{9.7cm} % For 1 column paper (peerreview)
  \else
    \begin{minipage}{8.7cm} % For 2 column paper (peerreview)
  \fi
\hrule \vspace{0.5em}%\centering
%   \begin{minipage}{13cm}    % For 1 coulumn paper
\hspace*{-1.1em}
  \ifonecol
    \begin{minipage}{9.5cm} % For 1 column paper (peerreview)
  \else
    \begin{minipage}{8.5cm} % For 2 column paper (peerreview)
  \fi
      {
        \footnotesize
        \def\baselinestretch{1}

        \begin{itemize}
        \item Initialization:

           \begin{enumerate}
             \item Let $\hsb_{0}$ be equal to the minimum \Ltwo\ norm solution of
             $\oureq$, obtained by pseudo-inverse of $\Ab$.

             \item Choose a suitable decreasing sequence for $\sigma$,
             $[\sigma_{1}\ldots\sigma_{J}]$ (see Remarks 5 and 6 of the text).
           \end{enumerate}

        \item For $j=1,\dots,J$: \vspace{0.5mm}\\
        \vrule\hspace{-1em}
        \begin{minipage}{8.2cm} 
          \begin{enumerate}
             \item \hspace{-0.8em} Let $\sigma=\sigma_i$.
             \item \hspace{-0.8em} Maximize (approximately) the function
             $F_{\sigma}$ on the feasible set $\Sm=\{\sbb \, | \, \Ab \sbb = \xb \}$ using $L$ iterations of the
             steepest ascent algorithm (followed by projection onto the feasible
             set):
             \begin{itemize}
               \item \hspace{-0.8em} Initialization: $\sbb=\hsb_{j-1}$.
               \item \hspace{-0.8em} For $\ell=1\dots L$ (loop $L$ times): \vspace{0.5mm} \\
                  \hspace*{-0.4em}\vrule\hspace{-1em}
                  \begin{minipage}{8cm} 
                    \begin{enumerate}
                       \item \hspace{-0.8em} Let
                       $\deltab\triangleq[s_{1}\exp{(-s_{1}^{2}/2\sigma^{2})},
                       \dots,s_{n}\exp{(-s_{n}^{2}/2\sigma^{2})}]^{T}$.

                       \item \hspace{-0.8em} Let $\sbb\leftarrow \sbb-\mu \deltab$ (where $\mu$ is a small positive constant).

                       \item \hspace{-0.8em} Project $\sbb$ back onto the feasible set $\Sm$:
                          \begin{equation*}
                             \sbb\leftarrow\sbb-\Ab^{T}(\Ab\Ab^{T})^{-1}(\Ab\sbb-\xb).
                          \end{equation*} %\vspace*{.05em}
                    \end{enumerate}
                  \end{minipage}\\\hspace*{-0.4em}\rule{2mm}{.5pt}
             \end{itemize}
             \item \hspace{-0.8em} Set $\hsb_{j}=\sbb$.\vspace{0.2em}
          \end{enumerate}
        \end{minipage}\\\rule{2mm}{.5pt}
        
        \item Final answer is $\hsb=\hsb_J$.
        \end{itemize}
      }
    \end{minipage}
    \vspace{1em} \hrule
  \end{minipage}\vrule \\
\caption{The final SL0 algorithm.} \label{fig: the alg}
\end{figure}

%% End of Algorithm %%%%%%%%%%%%%%%%%%%%%%%%%%%%%%%%%%%%%%%%%%%%%%%%%%%

{\bf Remark 1. \ } The internal loop (steepest ascent for a fixed
$\sigma$) is repeated a fixed and small number of times ($L$). In
other words, for increasing the speed, we do not wait for the
(internal loop of the) steepest ascent algorithm to converge. This
may be justified by the gradual decrease in the value of
$\sigma$, and the fact that for each value of $\sigma$, we do not
need the exact maximizer of $F_\sigma$. We just need to enter
the region near the (global) maximizer of $F_\sigma$ for
escaping from its local maximizers. See also Remarks~3 to~5 of
Section~\ref{sec: conv to sol}.

\vspace{0.5em}

{\bf Remark 2. \ } Steepest ascent consists of iterations of the
form $\sbb\leftarrow \sbb+\mu_j \nabla F_\sigma(\sbb)$. Here, the
step-size parameters $\mu_j$ should be decreasing, \ie for smaller
values of $\sigma$, smaller values of $\mu_j$ should be applied.
This is because for smaller values of $\sigma$, the function
$F_\sigma$ is more `fluctuating', and hence smaller step-sizes
should be used for its maximization. In fact, we may think about
changing the value of  $\sigma$ in (\ref{eq: Gaussian f}) and
(\ref{eq: Gaussian F}) as looking at the same curve (or surface)
at different `scales', where the scale is proportional to
$\sigma$. For having equal (\ie proportional) steps  of the
steepest ascent algorithm in these different scales, it is not
difficult to show\footnote{To see this, suppose that $s_1=r
\sigma_1$ in $F_{\sigma_1}$ corresponds to $s_2= r \sigma_2$ in
$F_{\sigma_2}$. Then $\mu_1 \nabla F_{\sigma_1} (s_1) / \mu_2
\nabla F_{\sigma_2} (s_2)=\sigma_1/\sigma_2$ results in
$\mu_1/\mu_2 = \sigma_1^2/\sigma_2^2$.} that $\mu_j$ should be
proportional to $\sigma^2$. Note that in Fig.~\ref{fig: the alg},
instead of  $\mu_j$, only a constant $\mu$ is appeared. The
reason is that by letting $\mu_j=\mu \sigma^2$ for some constant
$\mu$, we have $\sbb\leftarrow \sbb+ (\mu \sigma^2) \nabla \Fs =
\sbb-\mu \deltab$, where $\deltab \triangleq - \sigma^2 \,\nabla F_\sigma
=[s_{1}\exp{(-s_{1}^{2}/2\sigma^{2})},                       \dots,s_{n}\exp{(-s_{n}^{2}/2\sigma^{2})}]^{T}$.

\vspace{0.5em}

{\bf Remark 3. \ } According to the algorithm, each iteration
consists of an ascent step $s_i \leftarrow s_i-\mu s_i
\exp(-s_i^2/2\sigma^2), \, 1 \leq i \leq \Nu$, followed by a
projection step. If for some values of $i$ we have $|s_i|\gg\sigma$,
then the algorithm does not change the value of $s_i$ in that ascent
step; however it might be changed in the projection step. If we are
looking for a suitable large $\mu$ (to reduce the required number of
iterations), a suitable choice is to make the algorithm to force all
those values of $s_i$ satisfying $|s_i|\lesssim\sigma$ toward zero.
For this aim, we should have $\mu\exp(-s_i^2/2\sigma^2)\approx 1$,
and because $\exp(-s_i^2/2\sigma^2)\lesssim 1$ for
$|s_i|\lesssim\sigma$, the choice $\mu \gtrsim 1$ seems reasonable.
%In experimental results this choice is successfully utilized. @@@

\vspace{0.5em}

{\bf Remark 4. \ } The algorithm may work by initializing $\hsb_0$
(the initial estimation of the sparse solution) to an arbitrary
solution of $\oureq$. However, the discussion of
Section~\ref{sec: basic init} shows that the best initial value of
$\hsb_{0}$ is the minimum \Ltwo\ norm solution of $\oureq$, which
corresponds to $\sigma\to\infty$.
{\em In another point of view, one may think about the minimum
\Ltwo\ norm solution as a rough estimate of the sparse solution,
which will be modified in the future iterations of the
algorithm\/}. In fact, calculating minimum \Ltwo\ norm is one of
the earliest approaches used for estimating the sparsest solution
and is called the Method Of Frames (MOF)~\cite{ChenDS99}.

\vspace{0.5em}

{\bf Remark 5. \ } Having initiated the algorithm with the minimum
\Ltwo\ norm solution (which corresponds to $\sigma=\infty$), the
next value for $\sigma$ (\ie $\sigma_1$) may be chosen about two
to four times of the maximum absolute value of the obtained
sources ($\max_i |s_i|$). To see the reason, if we take for
example $\sigma
> 4 \max_i |s_i|$, then $\exp(-s_i^{2}/2\sigma^{2})>0.96\approx 1$
for all $1 \leq i \leq \Nu$, and comparison with (\ref{eq:
approximating prop 2}) shows that this value of $\sigma$ acts
virtually like infinity for all the values of $s_i$, $1 \leq i
\leq \Nu$ (the next remark, too, provides another reason through
another viewpoint to the algorithm).

For the next values of $\sigma$, we have used $\sigma_j=c\,
\sigma_{j-1}$, $j \geq 2$, where $c$ is usually chosen between 0.5
and 1. Its effect is experimentally studied in Section~\ref{sec:
experiments}).

\vspace{0.5em}

{\bf Remark 6. \ } Equation (\ref{eq: Gaussian F}) seems 
to simply count the ``inactive'' elements of $\sbb$. However,
instead of hard-thresholding ``$\mbox{inactive} \equiv |s_i| < \sigma \, ;
\mbox{ active} \equiv |s_i| > \sigma$'', criterion (\ref{eq: Gaussian F})
uses a soft-thresholding, for which $\sigma$ is the rough threshold.

{\bf Remark 7. \ } In applications where the zeros in the
sparsest $\sbb$ are exactly zero, $\sigma$ can be decreased
arbitrarily. In fact, in this case, its minimum value is
determined by the desired accuracy, as will be discussed in
Theorem~\ref{theo: solution convergence}. For applications in
which inactive elements of $\sbb$ are small but not exactly zero
(say that the `source' vector is noisy), the smallest $\sigma$
should be about one to two times of (a rough estimation of) the
standard deviation of this noise. This is because, while $\sigma$ is in this
range, (\ref{eq: approximating prop 2}) shows that the cost
function treats small (noisy) samples as zeros (\ie for which
$f_\sigma(s_i)\approx 1$). However, below this range, the
algorithm tries to `learn' these noisy values, and moves away
from the true answer (according to the previous remark, the
soft threshold should be such that all these noisy samples be considered
inactive). Restricting $\sigma_i$ to be above the
standard deviation of the noise, provides the robustness of this approach to
noisy sources (or mixtures), which was one of the difficulties in using the exact
\Lzero\ norm.

%%%%%%%%%%%%%%%%%%%%%%%%%%%%%%%%%%%%%%%%%%%%%%%%%%%%%%%%%%%%%%%%%%%
%%%%%%%%%%%                 New Section                 %%%%%%%%%%%
%%%%%%%%%%%%%%%%%%%%%%%%%%%%%%%%%%%%%%%%%%%%%%%%%%%%%%%%%%%%%%%%%%%
\section{Theoretical Analysis of the Algorithm} \label{sec: alg anal}

%%%%%%%%%%%%%%%%%%%%%%%%
%%%  New SubSection  %%%
%%%%%%%%%%%%%%%%%%%%%%%%
\subsection{Convergence Analysis} \label{sec: conv to sol}
In this section, we try to answer two questions for the noiseless case
(the noisy case will be considered in
Section~\ref{sec: Analysis Noisy}): a)~Does the basic idea of
Section~\ref{sec: Basic Principles} results in convergence to the
actual minimizer of the \Lzero\ norm (assumed to be unique
by~\cite{GoroR97,Dono04})?  and b)~If yes, how much should we
decrease $\sigma$ to achieve a desired accuracy?

%%%%%%%%%%%%%%%%%%%%%%%%%%%%%%%%%%%%%%%%%%%%%
% Deleted from the revision 1
%%%%%%%%%%%%%%%%%%%%%%%%%%%%%%%%%%%%%%%%%%%%%
Note that the algorithm of Fig.~\ref{fig: the alg} has two loops:
the external loop which corresponds to the basic ideas of
Section~\ref{sec: Basic Principles} for finding the sparsest
solution, and the internal loop which is a simple steepest ascent
algorithm for maximizing $F_\sigma(\sbb)$ for a fixed $\sigma$.
In the analysis of this section, 
it is assumed that the maximization of $F_\sigma(\sbb)$
has been exactly done for a fixed $\sigma$ (the maximization algorithm has not got trapped
into local maxima).
Note that we had proposed the gradual decrease in
$\sigma$ to escape from getting trapped into local maxima when
maximizing $F_\sigma$ for a fixed $\sigma$. 
A theoretical study to find the series $\sigma_j$, $j=1,\ldots J$, 
which guaranties the convergence 
is very tricky (if possible) and is not considered in this
paper. However, it will be experimentally addressed in the next section.

Assuming the maximization of $F_\sigma$ for fixed $\sigma$'s is
perfectly done, we show here that the estimation given by the
algorithm converges to the unique minimizer of the \Lzero\ norm.
In other words, we prove that the sequence of `global' maximizers
of $F_{\sigma}$'s will converge to the sparsest solution (which is
the basic idea of Section~\ref{sec: Basic Principles}), and try
to answer both above questions. 
%%%%%%%%%%%%%%%%%%%%%%%%%%%%%%%%%%%%%%%%%%%%%
% End of Deleted from the revision 1
%%%%%%%%%%%%%%%%%%%%%%%%%%%%%%%%%%%%%%%%%%%%%

Before stating the convergence theorem (Theorem~\ref{theo:
solution convergence}), we state three lemmas. Recall that
$\mathrm{null}(\Ab)=\{\sbb | \Ab \sbb = {\bf 0}\}$.

\vspace{0.5em}
\begin{lemma}\label{lem: s to zero}
Assume that the matrix $\Ab=[\ab_{1},\ab_{2},\cdots,\ab_{\Nu}] \in
\Rr^{\Ne \times \Nu}$ (where $\ab_{i}$ represents the $i$-th
column) has the property that all of its $\Ne \times \Ne$
sub-matrices are invertible, which is called the Unique Representation
Property (URP) in~\cite{GoroR97}\footnote{URP of $\Ab$ also guaranties 
that the sparsest solution is unique~\cite{GoroR97,Dono04}.}.
If $\Nu-\Ne$ elements of $\sbb \in \mathrm{null}(\Ab)$
converge to zero, then all of its elements (and hence $\sbb$)
will converge to zero, too.
\end{lemma}
\vspace{0.5em}

\begin{proof}
Without loss of generality, assume that all the columns of $\Ab$
are normalized, i.e. $\norm{\ab_{i}}{}=1$, $1 \leq i \leq \Nu$
(throughout the paper, $\norm{\cdot}{}$ stands for the \Ltwo\ or
Euclidean or Frobenius norm of a vector or matrix). Then, we have to show:
\begin{equation}\label{eq: lem to show}
  \begin{split}
  &\forall \beta>0, \quad \exists \alpha>0, \quad \mbox{such that } \forall \sbb \in
  \mathrm{null}(\Ab): \\
  & \mbox{$\Nu-\Ne$ elements of $\sbb$ have absolute values} \\& \qquad \qquad \qquad \mbox{less
  than
  $\alpha$}\Rightarrow \norm{\sbb}{} \leq \beta
  \end{split}
\end{equation}
Let $\sbb=(s_{1},s_{2},\cdots,s_{\Nu})^{T}$ be in
$\mathrm{null}(\Ab)$ and assume that the absolute values of at
least $\Nu-\Ne$ elements of it are smaller than $\alpha$. Let
$I_{\alpha}$ be the set of all indices $i$, for which
$|s_{i}|>\alpha$. Consequently, $|I_{\alpha}| \leq \Ne$, where
$|X|$ represents the cardinality (\ie number of elements) of a set
$X$. Then we write:
\begin{equation}
\begin{split}
\sum_{i=1}^{\Nu}{s_{i}\ab_{i}} = {\bf 0} \Rightarrow \sum_{i \in
I_{\alpha}}{s_{i}\ab_{i}} + \sum_{i \notin
I_{\alpha}}{s_{i}\ab_{i}}
= {\bf 0} \Rightarrow  \\
\norm{\sum_{i \in I_{\alpha}}{s_{i}\ab_{i}}} {} = \norm{\sum_{i
\notin I_{\alpha}}{s_{i}\ab_{i}}}{} \leq \sum_{i \notin
I_{\alpha}}{\norm{s_{i}\ab_{i}}{}} =  \\
\sum_{i \notin I_{\alpha}}{\underbrace{|s_{i}|}_{\leq
\alpha}\underbrace{\norm{\ab_{i}}{}}_{1}} \leq \sum_{i \notin
I_{\alpha}}{\alpha} = (\Nu - |I_{\alpha}|)\alpha \leq \Nu \alpha
\end{split}
\label{eq: start upper bound s}
\end{equation}

Let $\hAb$ be the sub-matrix of $\Ab$ containing only those
columns of $\Ab$ that are indexed by the elements of $I_{\alpha}$.
Thus $\hAb$ has at most $\Ne$ columns, and the columns of $\hAb$
are linearly independent, because of the URP of $\Ab$. Therefore,
there exists\footnote{Not that $\hAb$ is not necessarily a square
matrix and hence is not necessarily invertible. But it has
a left inverse, which is not necessarily unique. In this case
$\hAb^{-1}$ is just `one' of these inverses. For example, since 
$\hAb$ is tall and full-rank, its Moore-Penrose pseudoinverse is one of
these inverses.} a left inverse
$\hAb^{-1}$ for $\hAb$. Let $\bar{\sbb}$ and $\tsb$ denote those
sub-vectors of $\sbb$ which are, and which are not indexed by
$I_{\alpha}$, respectively. Then:
\begin{gather}
\begin{split}
\sum_{i \in I_{\alpha}}{s_{i}\ab_{i}}  = & \hAb \bar{\sbb}
\Rightarrow \norm{\bar{\sbb}}{}=\norm{(\hAb^{-1})(\sum_{i \in
I_{\alpha}}{s_{i}\ab_{i}})}{} \\
& \leq \norm{\hAb^{-1}}{}\cdot\norm{\sum_{i \in
I_{\alpha}}{s_{i}\ab_{i}}}{} \leq \norm{\hAb^{-1}}{}(\Nu \alpha)
\label{eq: th1 calc1}
\end{split}\\
\begin{split}
&\left.\begin{array}{l} \norm{\tsb}{} \leq \sum_{i \notin
I_{\alpha}}{|s_{i}|} \leq  (\Nu - |I_{\alpha}|) \alpha \leq
\Nu \alpha \\
\norm{\bar{\sbb}}{} \leq \norm{\hAb^{-1}}{} \Nu \alpha
\end{array} \right\} \Rightarrow \\
& \qquad \qquad\qquad\norm{\sbb}{} \leq \norm{\tsb}{} +
\norm{\bar{\sbb}}{} \leq (\norm{\hAb^{-1}}{}+1) \Nu \alpha
\label{eq: th1 calc2}
\end{split}
\end{gather}
Now, let $\mathcal{M}$ be the set of all submatrices $\hAb$ of
$\Ab$, consisting of at most $\Ne$ columns of $\Ab$. Then
$\mathcal{M}$ is clearly a finite set (in fact
$|\mathcal{M}|<2^\Nu$). Let\footnote{Note that
the calculation of $M$ is difficult in the cases where $\Nu$ and
$\Ne$ are large. Calculation of the exact value of $M$ requires a
computation complexity larger than $\binom{\Nu}{\Ne}$ which can be
impractical for large values of $\Nu$ and $\Ne$.}
\begin{equation}
M=\max\{ \norm{\hAb^{-1}} \,\,\,\, | \, \hAb \in \mathcal{M}
\},
\label{eq: M def}
\end{equation}
then
\begin{equation}
\norm{\sbb}{} \leq (\norm{\hAb^{-1}}{}+1)\Nu\alpha \leq
(M+1)\Nu\alpha.
\label{eq: Upper Bound s}
\end{equation}
$M$ is a constant and its value depends only on the matrix $\Ab$.
Therefore, for each $\beta$ it suffices to choose
$\alpha=\beta/\Nu (M+1)$.
\end{proof}

The above proof (calculations~(\ref{eq: start upper bound s}) to~(\ref{eq: Upper Bound s})) results also in the following corollary:

\medskip

\begin{coro}\label{coro: main coro}
If $\Ab\in\Rr^{\Ne \times \Nu}$ satisfies URP, and $\sbb\in\mathrm{null}(\Ab)$ has at most $n$ elements with absolute values greater than $\alpha$, then $\|\sbb\|<(M+1)\Nu\alpha$, where $M$ is as defined in (\ref{eq: M def}).
\end{coro}

\medskip

\begin{lemma}\label{lem: distance to true solution}
Let a function $\fs(s)$ have the properties $\fs(0)=1$ and $\forall s, 0\le \fs(s) \le 1$, and let $\Fs(\sbb)$ be defined as in (\ref{eq: Gaussian F}). Assume $\Ab$ satisfies the URP, and 
let $\Sm \triangleq \{\sbb \, | \, \Ab \sbb = \xb \}$. Assume that there exists a (sparse) solution $\sbb^{0} \in \Sm$ for which $\norm{\sbb^{0}}{0}=k \leq \Ne/2$ (such a sparse solution is unique~\cite{GoroR97,Dono04}). Then, if for a solution $\hsb=(\hat s_1,\dots, \hat s_m)^T \in \Sm$:
\begin{equation}
\Fs(\hsb) \ge m - (n-k),
\label{eq: limit condition}
\end{equation}
and if $\alpha>0$ is chosen such that the $\hat s_i$'s with absolute values greater than $\alpha$ satisfy $\fs(\hat s_i) \le \frac{1}{m}$, then:
\begin{equation}
\| \hsb -\sbb^0 \| < (M+1)m \alpha,
\label{eq: upper bound 1}
\end{equation}
where $M$ is as defined in (\ref{eq: M def}).
\end{lemma}

\begin{proof}
Let $I_\alpha$ be the set of all indices $i$ for which $|\hat s_i| > \alpha$, 
and denote its number of elements by $|I_\alpha|$. Then:
\begin{equation*}
\begin{split}
\Fs(\hsb) &= \sum_{i=1}^{\Nu}
\fs(\hat s_i) \\
&= 
\underbrace{\sum_{i \notin I_\alpha} \underbrace{\fs(\hat s_i)}_{\leq 1}}_{\le m-|I_\alpha|}
+
\underbrace{\sum_{i \in I_{\sigma}} \underbrace{\fs(\hat s_i)}_{< \frac{1}{m}}}_{< m \cdot \frac{1}{m}=1}
< m - |I_\alpha| + 1.
%\Nu-|I_{\sigma}|+ \sum_{i \in I_{\sigma}} \fs(s^{\sigma}_i)
\end{split}
\end{equation*}
Combining this result with (\ref{eq: limit condition}), we obtain: 
\begin{equation*}
\begin{split}
m-(n-k) \le \Fs(\hsb) < m - |I_\alpha| + 1 \\ \Rightarrow
|I_\alpha| < n-k + 1 \Rightarrow |I_\alpha| \le n-k.
\end{split}
\end{equation*}
Consequently, at most $n-k$ elements of $\hsb$ have absolute values greater than $\alpha$. Since $\sbb^0$ has exactly $k$ non-zero elements, we conclude that $\hsb-\sbb^0$ has at most $(n-k)+k=n$ elements with absolute values greater than $\alpha$. Moreover, $(\hsb-\sbb^0)\in \mathrm{null}(\Ab)$
(because $\Ab(\hsb-\sbb^0)=\xb - \xb = {\bf 0}$), and hence Corollary~\ref{coro: main coro} implies (\ref{eq: upper bound 1}).
\end{proof}

\medskip

\begin{coro}\label{coro: Guasian upper bound}
For the Gaussian family (\ref{eq: Gaussian f}), if (\ref{eq: limit condition}) holds for a solution $\hsb$, then:
\begin{equation}
\| \hsb -\sbb^0 \| < (M+1)m \sigma \sqrt{2 \ln m}\, .
\label{eq: upper bound Gaussian}
\end{equation}
\end{coro}

\medskip

\begin{proof}
For Gaussian family (\ref{eq: Gaussian f}), the $\alpha$ of the above lemma can be chosen as $\alpha = \sigma \sqrt{2 \ln m}$, because for $|\hat s_i| > \sigma \sqrt{2 \ln m}$:
\begin{equation*}
	\fs({\hat s_i})=
	\exp\left\{-\frac{\hat s_i^2}{2 \sigma^2}\right\} < 
	\exp\left\{-\frac{\sigma^2 \cdot 2 \ln m}{2 \sigma^2}\right\} = \frac{1}{m}.
\end{equation*}
Moreover, this family satisfies the other conditions of the lemma.
\end{proof}

\medskip

\begin{lemma}\label{lem: max satisfies lim cond}
Let $\fs$, $\Fs$, $\Sm$ and $\sbb^0$ be as in Lemma~\ref{lem: distance to true solution}, and let $\sbb^\sigma$ be the maximizer of $F_\sigma(\sbb)$ on $\Sm$. Then $\sbb^\sigma$ satisfies (\ref{eq: limit condition}).
\end{lemma}

\begin{proof}
We write:
\begin{align}
\Fs(\sbb^\sigma) & \ge \Fs(\sbb^0) && \mbox{(because $\sbb^\sigma$ is the maximizer)} \nonumber \\
& \ge m-k  &&\mbox{(see below)} \label{eq: max s limit} \\
& \ge m - (n-k) &&\mbox{(because $k \le \frac{n}{2}$).} \nonumber
\end{align}
The second inequality was written because $\sbb^0$ has $m-k$ zeros, and hence in the summation (\ref{eq: Gaussian F}) there are $m-k$ ones, and the other terms are non-negative.
\end{proof}

\medskip

Note that Lemma~\ref{lem: max satisfies lim cond} and Corollary~\ref{coro: Guasian upper bound} prove together that for the Gaussian family (\ref{eq: Gaussian f}), $\argmax_{\Ab \sbb = \xb} \Fs(\sbb) \to \sbb^0$ as $\sigma \to 0$. This result can, however, be stated for a larger class of functions $\fs$, as done in the following Theorem.

\medskip 

\begin{theorem}
\label{theo: solution convergence} Consider a family of univariate
functions $\fs$, indexed by $\sigma$, $\sigma \in \Rr^{+}$,
satisfying the set of conditions:
\begin{enumerate}
\item $\lim_{\sigma \to 0} \fs(s)=0 \qquad \textrm{; for all }s \neq 0$
\item ${\fs(0)}=1 \qquad \textrm{; for all }\sigma \in \Rr^{+}$
\item $0 \leq \fs(s) \leq 1 \qquad \textrm{; for all }\sigma \in \Rr^{+}, s \in \Rr$
\item For each positive values of $\nu$ and $\alpha$, there exists
$\sigma_{0} \in \Rr^{+}$ that satisfies: \label{item: cond4}
\begin{equation}
|s|>\alpha \Rightarrow \fs(s)<\nu \quad \textrm{; for all
}\sigma<\sigma_{0}.
\label{eq: cond 4 main theorem}
\end{equation}
\end{enumerate}
Assume $\Ab$ satisfies the URP, and let $\Sm$, $\Fs$ and $\sbb^0$
be as defined in Lemma~\ref{lem: distance to true solution}, and 
$\sbb^\sigma=(s^\sigma_1, \dots, s^\sigma_m)^T$ be the maximizer of $F_\sigma(\sbb)$ on $\Sm$. Then:
\begin{equation}
\lim_{\sigma \to 0}{\sbb^\sigma}=\sbb^{0}.
\label{eq: s-sig to s0}
\end{equation}
\end{theorem}

\medskip

\begin{proof}
To prove (\ref{eq: s-sig to s0}), we
have to show that:
\begin{equation}\label{eq: theo 1 to prove}
  \forall \beta>0 \quad \exists \sigma_0>0, \quad \forall \sigma<\sigma_0 \quad \norm{\sbb^{\sigma}-\sbb^0}{}<\beta.
\end{equation}
For each $\beta$, let $\alpha=\beta/\Nu (M+1)$, where $M$ is as defined in (\ref{eq: M def}). Then for this $\alpha$ and $\nu=\frac{1}{m}$, condition~\ref{item: cond4} of the theorem gives a $\sigma_0$ for which (\ref{eq: cond 4 main theorem}) holds. We show that this is the $\sigma_0$ we were seeking for in (\ref{eq: theo 1 to prove}). Note that $\forall \sigma < \sigma_0$, (\ref{eq: cond 4 main theorem}) states that for $s^\sigma_i$ 's with absolute values greater than $\alpha$ we have $\fs(s^\sigma_i)<\frac{1}{m}$. Moreover, Lemma~\ref{lem: max satisfies lim cond} states that $\sbb^\sigma$ satisfies~(\ref{eq: limit condition}). Consequently, all the conditions of Lemma~\ref{lem: distance to true solution} have been satisfied, and hence it implies that
$\| \sbb^\sigma -\sbb^0 \| < (M+1)m \alpha = \beta$.	
\end{proof}

\medskip 

{\bf Remark 1. \ } The Gaussian family (\ref{eq: Gaussian f})
satisfies conditions 1 through 4 of Theorem \ref{theo: solution
convergence}. In fact, conditions 1, 2 and 3 are obvious. To see
condition 4, it is sufficient to choose
$\sigma_{0}^{2}=-\alpha^{2}/(2\ln{\nu})$ if $\nu<1$, or to choose
any arbitrary $\sigma_{0}^{2}\in \Rr^+$ if $\nu\geq1$.
Families of functions defined by (\ref{eq: trig f}), (\ref{eq:
hyperbolic f}) and (\ref{eq: second order mobius f}) also satisfy
the conditions of this theorem.

\vspace{0.5em}

{\bf Remark 2. \ } Using Corollary~\ref{coro: Guasian upper bound},
where using Gaussian family~(\ref{eq: Gaussian f}), to ensure an arbitrary accuracy $\beta$ in estimation of the
sparse solution $\sbb^0$, it suffices to choose:
\begin{displaymath}
\sigma<\frac{\beta}{\Nu\sqrt{2\ln{\Nu}}(M+1)},
\end{displaymath}
and do the optimization of $\Fs$ subject to $\oureq$. 

\vspace{0.5em}

{\bf Remark 3. \ } Consider the set of solutions $\hsb^\sigma$ in
$\Sm$, which might not be the absolute maxima of functions $\Fs$
on $\Sm$, but satisfy the condition
\begin{equation}
\Fs(\hsb^\sigma) \geq \Nu-(\Ne-k).
\label{eq: limit f for SL0 without L}
\end{equation}
By following a similar approach to the proof of Theorem \ref{theo:
solution convergence}, it can be proved that $\lim_{\sigma \to
0}\hsb^\sigma = \sbb^{0}$. In other words, for the steepest ascent
of the internal loop, it is not necessary to reach the absolute
maximum. It is just required to achieve a solution in which $\Fs$
is large enough (see also Remark~1 of Section~\ref{sec: alg}).
%Clearly, the maximizers $\sbb^{\sigma}$ satisfy the condition.

\vspace{0.5em}

{\bf Remark 4. \ } The previous remark proposes another version 
of SL0 in which there is no need to set a parameter $L$: 
Repeat the internal loop of Fig.~\ref{fig: the alg}
until $F_\sigma(\sbb)$ exceeds $m-n/2$ (the worst case of the limit given
by~(\ref{eq: limit f for SL0 without L})) or $m-(n-k)$ if $k$ is known a priori (note that (\ref{eq: max s limit}) 
implies the maximizer of $F_\sigma(\sbb)$ for a fixed $\sigma$ surely exceeds both of these limits). 
The advantage of such a version is that
if it converges, then it is guaranteed that the estimation error is bounded
as in (\ref{eq: upper bound Gaussian}), in which $\sigma$ is replaced with $\sigJ$. 
It has however two disadvantages:
firstly, it slows down the algorithm
because exceeding the limit $m-(n-k)$ for each $\sigma$ is not necessary (it is just sufficient);
and secondly, because of the possibility that
the algorithm runs into an infinite
loop because $F_\sigma(\sbb)$ cannot exceed this limit (this occurs if the chosen sequence of $\sigma$ has not been resulted in escaping from local maxima).

\vspace{0.5em}

{\bf Remark 5. \ } As another consequence, Lemma~\ref{lem: s to zero} 
provides an upper bound on the estimation error $\|\hsb-\sbb^0\|$, only
by having an estimation $\hsb$ (which satisfies $\Ab \hsb = \xb$):
Begin by sorting the elements of $\hsb$ in 
descending order and let $\alpha$
be the absolute value of the ($\left\lfloor \frac{n}{2} \right\rfloor + 1$)'th element.
Since $\sbb^0$ has at most $n/2$ non-zero elements, $\hsb-\sbb^0$ has at most $n$ 
elements with absolute values greater than $\alpha$. Moreover, $(\hsb-\sbb^0)\in\mathrm{null}(\Ab)$ and 
hence Corollary~\ref{coro: main coro} implies that
$\|\hsb-\sbb^0\| \le (M+1)m \alpha$, where $M$ is as defined in
(\ref{eq: M def}). This result is consistent with the heuristic ``if $\hsb$ has at most
$n/2$ `large' components, the uniqueness of the sparsest solution insures that $\hsb$
is close to the true solution''.

\subsection{Relation to minimum norm 2 solution} \label{sec: exact relation min norm 2}

In section \ref{sec: basic init}, it was stated and informally
justified (for the Gaussian family~(\ref{eq: Gaussian f})) that
for very large $\sigma$'s, the maximizer of the function $\Fs$
subject to $\oursystem$ is the minimum $\ell^2$-norm solution of
$\oursystem$. This result can be more accurately proved, and also
generalized to a wider class of functions:

\vspace{0.5em}
\begin{theorem} \label{theo: min L2 norm}
Consider a family of one variable functions $f_{\sigma}(\cdot)$,
parameterized by $\sigma \in \mathbb{R}^{+}$, satisfying the
set of conditions:
\begin{enumerate}
\item All functions $f_{\sigma}$ are scaled versions of some
analytical function $f$, that is, $f_{\sigma}(s)=f(s/\sigma)$
\item $\forall s \in \mathbb{R}, \quad 0 \leq f(s) \leq 1$
\item $f(s)=1 \Leftrightarrow s=0$
\item $f'(0)=0$
\item $f''(0)<0$
\end{enumerate}

Assume that the matrix $\Ab$ is full-rank and let
$\hsb\triangleq\argmin_{\oureq}{\norm{\sbb}{}}=\Ab^{T}{(\Ab\Ab^{T})}^{-1}\xb$
be the minimum $\ell^{2}$-norm solution of the USLE $\oursystem$.
Then:
\begin{displaymath}
\lim_{\sigma \to \infty}{\argmax_{\oureq} F_{\sigma}{(\sbb)}}=\hsb.
\end{displaymath}
\end{theorem}

\vspace{0.5em}

\begin{proof}
Let $\sbb^\sigma=(s^\sigma_1, \dots, s^\sigma_m)^T=\argmax_{\oureq} F_{\sigma}{(\sbb)}$. Then,
we have to show that $\lim_{\sigma \to \infty}
\sbb^\sigma=\hsb=(\hat s_1,\dots, \hat s_\Nu)^T$.

First we show that:
\begin{equation}
\lim_{\sigma \to \infty} \frac{\sbb^\sigma}{\sigma}={\bf 0}.
 \label{eq: theo rel norm 2 first to prove}
\end{equation}
Since $\sbb^{\sigma}$ is the maximizer of $F_{\sigma}$, we have:
\begin{equation}\label{eq: theo rel norm 2 temp 11}
  F_{\sigma}(\sbb^{\sigma}) \geq F_{\sigma}{(\hsb)},
\end{equation}
and hence:
\begin{gather}
\lim_{\sigma \to \infty}F_{\sigma}(\sbb^{\sigma}) \geq
\lim_{\sigma \to \infty}F_{\sigma}(\hsb)
=\sum_{i=1}^{\Nu}\lim_{\sigma \to \infty} f(\hat{s}_i/\sigma)=\Nu
\nonumber \\
\Rightarrow \sum_{i=1}^{\Nu}\lim_{\sigma \to \infty}
f(s^{\sigma}_i/\sigma)=F_{\sigma}(\sbb^{\sigma}) \geq \Nu.
\label{eq: theo rel norm 2 temp1}
\end{gather}
On the other hand, assumption 2 implies that for all $1 \leq i
\leq \Nu$, $0 \leq \lim_{\sigma \to \infty}
f(s^{\sigma}_i/\sigma) \leq 1$. Combining this with (\ref{eq:
theo rel norm 2 temp1}), we have:
\begin{equation}
  \lim_{\sigma \to \infty} f(s^{\sigma}_i/\sigma)=1 \quad \textrm{; for } 1 \leq i \leq \Nu.
\end{equation}
This result, combined with assumption 3 (that is, $f^{-1}(1)=0$) and
the continuity of $f$ implies that for all $1 \leq i \leq \Nu$,
$\lim_{\sigma \to \infty} s^{\sigma}_i/\sigma=0$; from which
(\ref{eq: theo rel norm 2 first to prove}) is deducted.

Now, let $\gamma=\frac{-1}{2}f''(0)>0$. Then we can write
\begin{displaymath}
f(s)=1-\gamma s^2+g(s),
\end{displaymath}
where:
\begin{equation} \label{eq: g order}
\lim_{s \to 0}{\frac{g(s)}{s^2}}=0.
\end{equation}
Then:
\begin{displaymath}
F_{\sigma}(\sbb)=\Nu-\frac{\gamma}{\sigma^2}\sum_{i=1}^{\Nu}{s_{i}^{2}}+\sum_{i=1}^{\Nu}{g(s_i/\sigma)}.
\end{displaymath}
Consequently, (\ref{eq: theo rel norm 2 temp 11}) can be written as:
\begin{gather}
\frac{\gamma}{\sigma^2}\sum_{i=1}^{\Nu}(s^{\sigma}_i)^2-\sum_{i=1}^{\Nu}
g(s^{\sigma}_i/\sigma) \leq
\frac{\gamma}{\sigma^2}\sum_{i=1}^{\Nu}(\hat{s}_i)^2-\sum_{i=1}^{\Nu}
g(\hat{s}_i/\sigma)
\nonumber \\
\begin{split}
\Rightarrow  \norm{&\sbb^{\sigma}}{}^{2}-\norm{\hsb}{}^{2} \leq
\frac{\sigma^2}{\gamma}\sum_{i=1}^{\Nu}g(s^{\sigma}_{i}/\sigma)-
\frac{\sigma^2}{\gamma}\sum_{i=1}^{\Nu}g(\hat{s}_{i}/\sigma)\nonumber \\
&=\frac{1}{\gamma}\sum_{i=1}^{\Nu}\frac{g(s^{\sigma}_{i}/\sigma)}{(s^{\sigma}_{i}/\sigma)^2}(s^{\sigma}_{i})^2-
\frac{1}{\gamma}\sum_{i=1}^{\Nu}\frac{g(\hat{s}_{i}/\sigma)}{(\hat{s}_{i}/\sigma)^2}(\hat{s}_{i})^2
 \nonumber \\
&\leq\frac{1}{\gamma}|\sum_{i=1}^{\Nu}\frac{g(s^{\sigma}_{i}/\sigma)}{(s^{\sigma}_{i}/\sigma)^2}(s^{\sigma}_{i})^2|+
\frac{1}{\gamma}|\sum_{i=1}^{\Nu}\frac{g(\hat{s}_{i}/\sigma)}{(\hat{s}_{i}/\sigma)^2}(\hat{s}_{i})^2|
 \nonumber \\
&\leq\frac{1}{\gamma}(\sum_{i=1}^{\Nu}|\frac{g(s^{\sigma}_{i}/\sigma)}{(s^{\sigma}_{i}/\sigma)^2}|)
\norm{\sbb^{\sigma}}{}^2+
\frac{1}{\gamma}(\sum_{i=1}^{\Nu}|\frac{g(\hat{s}_{i}/\sigma)}{(\hat{s}_{i}/\sigma)^2}|)
\norm{\hsb}{}^2, \nonumber
\end{split}
\end{gather}
where for the last inequality, we have used the inequality:
\begin{displaymath}
|\sum_{i \in I,j \in J}x_i y_j| \leq \sum_{i \in I}|x_i|\sum_{j
\in J}|y_j|.
\end{displaymath}
Finally:
\begin{gather}
\norm{\sbb^{\sigma}}{}^{2} \leq \norm{\hsb}{}^2
\frac{1+\frac{1}{\gamma}(\sum_{i=1}^{\Nu}|\frac{g(\hat{s}_{i}/\sigma)}{(\hat{s}_{i}/\sigma)^2}|)}
{|1-\frac{1}{\gamma}(\sum_{i=1}^{\Nu}|\frac{g(s^{\sigma}_{i}/\sigma)}{(s^{\sigma}_{i}/\sigma)^2}|)|}, \nonumber \\
\lim_{\sigma \to \infty} \hat{s}_{i}/\sigma =0 \Rightarrow \lim_{\sigma \to \infty} \frac{g(\hat{s}_{i}/\sigma)}{(\hat{s}_{i}/\sigma)^2}=0 \quad \textrm{(from (\ref{eq: g order}))},\nonumber \\
\lim_{\sigma \to \infty} s^{\sigma}_{i}/\sigma =0 \Rightarrow
\lim_{\sigma \to \infty}
\frac{g(s^{\sigma}_{i}/\sigma)}{(s^{\sigma}_{i}/\sigma)^2}=0 \quad
\textrm{(from %(\ref{eq: theo rel norm 2 first to prove}) and 
(\ref{eq: g order}))}, \nonumber \\
\Rightarrow \lim_{\sigma \to \infty} \norm{\sbb^{\sigma}}{}^2 \leq
\norm{\hsb}{}^2. \label{eq: norm 2 relation}
\end{gather}
Noting that $\hsb$ is the minimum $\ell^2$-norm solution of
$\oursystem$, $\norm{\sbb^{\sigma}}{}^2 \geq \norm{\hsb}{}^2$, and
hence $\lim_{\sigma \to \infty} \norm{\sbb^{\sigma}}{}^2 \geq
\norm{\hsb}{}^2$. Combining this with (\ref{eq: norm 2 relation}),
we have:
\begin{equation}\label{eq: theo 2 rel norm 2 temp 34}
\lim_{\sigma \to \infty} \norm{\sbb^{\sigma}}{}^2 =
\norm{\hsb}{}^2.
\end{equation}
On the other hand, since $\hsb$ is the minimum $\ell^{2}$-norm
solution of $\oursystem$, it is perpendicular to any vector
contained in $\mathrm{null}(\Ab)$. This is because $\forall \vb
\in \mathrm{null}(\Ab), \Ab\vb={\bf 0}$, and hence
$\vb^T\hsb=\vb^T\Ab^T(\Ab\Ab^T)^{-1}\xb=
(\Ab\vb)^T(\Ab\Ab^T)^{-1}\xb={\bf 0}$.
Consequently, $\hsb$ is perpendicular to $\sbb^{\sigma}-\hsb$.
Therefore:
\begin{gather}
\norm{\sbb^{\sigma}}{}^2=\norm{\hsb}{}^2+\norm{\sbb^{\sigma}-\hsb}{}^2 \nonumber \\
\Rightarrow \lim_{\sigma \to \infty}
\norm{\sbb^{\sigma}}{}^2=\norm{\hsb}{}^2+\lim_{\sigma \to \infty}
\norm{\sbb^{\sigma}-\hsb}{}^2. \nonumber
\end{gather}
Combining this with (\ref{eq: theo 2 rel norm 2 temp 34}) we have
$\lim_{\sigma \to \infty} \norm{\sbb^{\sigma}-\hsb}{}^2=0$, and
hence $\lim_{\sigma \to \infty} \sbb^{\sigma} = \hsb$.
\end{proof}

\vspace{0.5em}

{\bf Remark 1. \ } The Gaussian family (\ref{eq: Gaussian f})
satisfies the conditions 1 through 5 of Theorem \ref{theo: min L2
norm}. Therefore, for this family of functions, the minimum
$\ell^2$-norm solution is the optimal initialization. Family of
functions defined by (\ref{eq: second order mobius f}) also
satisfies the conditions of this theorem, contrary to those defined in
(\ref{eq: trig f}) and (\ref{eq: hyperbolic f}) which are not analytic.

%%%%%%%%%%%%%%%%%%%%%%%%
%%%  New SubSection  %%%
%%%%%%%%%%%%%%%%%%%%%%%%
\subsection{The noisy case}  % About final value of sigma in no-noise and noisy cases
\label{sec: Analysis Noisy}

As shown in the proof of Theorem~\ref{theo: solution
convergence}, in the noiseless case, a smaller value of $\sigma$
results in a more accurate solution and it is possible to achieve
solutions as accurate as desired by choosing small enough values
of $\sigma$. However, this is not the case in the presence of
additive noise\footnote{The `noise' in this context has two meanings:
1)~the noise in the source vector $\sbb$ means that the inactive
elements of $\sbb$ are not exactly equal to zero; and 2)~the
(additive) noise in the sensors means that $\xb$ is not exactly equal to
$\Ab \sbb$. In the theorems of this section, only the second type
of noise has been considered, and it is assumed that the first
type does not exist. In other words, the inactive elements of
$\sbb$ are assumed to be exactly zero.}, that is, if $\xb =
\Ab \sbb + \nb$. In fact, noise power bounds maximum achievable
accuracy. We state a theorem in this section, which can be
considered as an extension of Theorem \ref{theo: solution
convergence} to the noisy case.

First, we state the following lemma, which can be considered as a
generalization to Lemma~\ref{lem: s to zero}.

\vspace{0.5em}

\begin{lemma} \label{lem: noisy}
Let $\Ab$ satisfy the conditions of Lemma~\ref{lem: s to zero},
and assume that the vector $\sbb$ has $\Nu-\Ne$ elements with
absolute values less than $\alpha$, and
$\norm{\Ab\sbb}{}<\epsilon$. Then $\norm{\sbb}{}<\beta$, where
\begin{displaymath}
\beta=(M+1)(\Nu\alpha+\epsilon),
\end{displaymath}
and $M$ is as defined in (\ref{eq: M def}).
\end{lemma}

\vspace{0.5em}

Note that in this lemma, instead of condition $\Ab\sbb=\bf{0}$,
we have a relaxed condition $\norm{\Ab\sbb}{}<\epsilon$.
Lemma~\ref{lem: s to zero} is the special (noiseless) case of this lemma where
$\epsilon \to 0$.

\begin{proof}
Let $I_{\alpha}$, $\hAb$, $\tsb$, $\bsb$ and $M$ be defined as
in the proof of Lemma~\ref{lem: s to zero}. Then
\begin{gather}
\norm{\sum_{i=1}^{\Nu}{s_{i}\ab_{i}}}{} < \epsilon \Rightarrow
\norm{\sum_{i \in I_{\alpha}}{s_{i}\ab_{i}} + \sum_{i \notin
I_{\alpha}}{s_{i}\ab_{i}}}{}
< \epsilon \Rightarrow \nonumber \\
\norm{\sum_{i \in I_{\alpha}}{s_{i}\ab_{i}}} {} < \norm{\sum_{i
\notin I_{\alpha}}{s_{i}\ab_{i}}}{} + \epsilon \leq \sum_{i \notin
I_{\alpha}}{\norm{s_{i}\ab_{i}}{}} + \epsilon = \nonumber \\
\sum_{i \notin I_{\alpha}}{|s_{i}|\norm{\ab_{i}}{}} + \epsilon \leq
\sum_{i \notin I_{\alpha}}{\alpha} + \epsilon  =  (\Nu -
|I_{\alpha}|)\alpha + \epsilon \leq \Nu \alpha + \epsilon. \nonumber
\end{gather}
Therefore, by repeating the calculations of (\ref{eq: th1 calc1})
and (\ref{eq: th1 calc2}), we obtain $\norm{\sbb}{}<(M+1)
(\Nu\alpha + \epsilon)$.
\end{proof}

\vspace{0.5em}
\begin{theorem}
\label{theo: noisy case} Let $\Sm_\epsilon=\{\sbb  | \, \,
\norm{\Ab\sbb-\xb}{} < \epsilon\}$, where $\epsilon$ is an
arbitrary positive number, and assume that the matrix $\Ab$ and
functions $\fs$ satisfy the conditions of Theorem \ref{theo:
solution convergence}. Let $\sbb^0 \in \Sm_\epsilon$ be a sparse
solution, and assume that $\fs$ satisfies the extra conditions:
\begin{enumerate}
\item There exists $\gamma>0$ such that
\begin{displaymath}
%\label{eq: noisy condition}
|\frac{d}{ds}\fs(s)|<\gamma/\sigma \quad \textrm{; for all $\sigma>0$
and all $s$}
\end{displaymath}
\item For each positive values of $\nu$ and $\sigma_{0}$, there
exists an $\alpha>0$ that satisfies:\\
\begin{displaymath}
|s|>\alpha \Rightarrow \fs(s)<\nu \quad \textrm{; for all
}\sigma<\sigma_{0}
\end{displaymath}
\end{enumerate}
Let $M$ and $k$ be defined as in Theorem \ref{theo:
solution convergence}. Then under the condition $k<\Ne/2$, by
choosing
\begin{equation}
\label{eq: choose sigma noisy}
\sigma_0=\frac{\Nu\gamma\epsilon\norm{\Ab^T(\Ab\Ab^T)^{-1}}{}}{(\Ne-2k)},
\end{equation}
and optimizing $F_{\sigma_0}$, the sparse solution can be
estimated with an error smaller than
\begin{displaymath}
(M+1) (\Nu\alpha + \epsilon),
\end{displaymath}
where $\alpha$ is the value for which the condition 2 holds for
$\sigma_0$ and $\nu=1/\Nu$.
\end{theorem}

\vspace{0.5em}

\begin{proof}
Let $\nb\triangleq\Ab\sbb^0-\xb$. Then, $\sbb^0 \in \Sm_\epsilon$
means that $\norm{\nb}{}<\epsilon$. By defining
$\tnb\triangleq\Ab^T(\Ab\Ab^T)^{-1}\nb$, we have:
\begin{displaymath}
\xb=\Ab\sbb^0+\nb=\Ab\sbb^0+\Ab\tnb=\Ab(\sbb^0+\tnb)=\Ab\tsb,
\end{displaymath}
where $\tsb \triangleq \sbb^0+\tnb$. Let $\sbb^\sigma$ be the
maximizer\footnote{Note that, $\sbb^\sigma$ is not necessarily
maximizer of $\Fs$ on the whole $\Sm_\epsilon$.} of $\Fs$ on
$\oureq$, as defined in Theorem \ref{theo: solution convergence}.
When working with $\ell^0$-norm, no matter how much small is
$\epsilon$ and how much sparse is $\sbb^0$, $\tsb$ is not
necessarily sparse. However, as will be discussed, because $\Fs$
is continuous and $\norm{\nb}{}$ is small, the value of $\Fs$ at
$\tsb$ is close to its value at $\sbb^0$ (and thus, is large). In
fact:
\begin{equation*}
  \Fs(\tsb)=\Fs(\sbb^0+\tnb)\simeq \Fs(\sbb^0)+\nabla \Fs(\sbb^0)\cdot\tnb.
\end{equation*}
By defining $g(t)\triangleq \Fs(\sbb^0+\tnb t)$, we have
$g(0)=\Fs(\sbb^0)$ and $g(1)=\Fs(\sbb^0+\tnb)=\Fs(\tsb)$. Using
the mean value theorem, there exists a $0 \leq t \leq 1$ such
that:
\begin{equation*}
  \begin{split}
    |\Fs(\tsb)-\Fs(\sbb^0)|=|g(1)-g(0)| \leq (1-0)g'(t) \\
    =\nabla\Fs(\sbb^0+\tnb t) \cdot \tnb \leq
    \norm{\nabla\Fs(\sbb^0+\tnb t)}{} \cdot \norm{\tnb}{}
  \end{split}
\end{equation*}
We write:
\begin{gather}
\left\{ \begin{array}{l} \forall s \quad |\frac{d}{ds}\fs(s)|  <
\gamma/\sigma \Rightarrow
\norm{\nabla\Fs(\sbb^0+\tnb t)}{} < \Nu\gamma/\sigma \\
\norm{\tnb}{}=\norm{\Ab^T(\Ab\Ab^T)^{-1}\nb}{} <
\norm{\Ab^T(\Ab\Ab^T)^{-1}}{}\epsilon \end{array} \right\}
\Rightarrow
\nonumber \\
|\Fs(\tsb)-\Fs(\sbb^0)|<\Nu\gamma\epsilon\norm{\Ab^T(\Ab\Ab^T)^{-1}}{}/\sigma
\nonumber
\end{gather}
Let choose $\sigma_0$ according to (\ref{eq: choose sigma noisy}).
Then:
\begin{eqnarray}
\left\{ \begin{array}{l}
|F_{\sigma_0}(\tsb)-F_{\sigma_0}(\sbb^0)|<\Ne-2k \\
F_{\sigma_0}(\sbb^0) \geq \Nu-k
\end{array} \right. \Rightarrow F_{\sigma_0}(\tsb)> \Nu - (\Ne-k)
\nonumber
\end{eqnarray}
The vector $\sbb^0$ does not necessarily satisfy $\oureq$, however
we have chosen $\tsb$ to be the projection of $\sbb^0$ onto the
subspace $\oureq$. Hence, $\tsb$ satisfies $\oureq$ and since
$\sbb^{\sigma_0}$ is the maximizer of $F_{\sigma_0}$ on $\oureq$,
$F_{\sigma_0}(\sbb^{\sigma_0})>\Nu-(\Ne-k)$. Consequently, by
choosing $\alpha$ as the value for which the condition 2 holds for
$\nu=1/\Nu$ and $\sigma_0$, and following the same steps as in
the proof of Theorem \ref{theo: solution convergence}, we
conclude that at most $\Ne-k$ elements of $\sbb^{\sigma_0}$ can
have absolute values greater than $\alpha$. Then, since $\sbb^0$
has at most $k$ nonzero elements, $(\sbb^0-\sbb^{\sigma_0})$ has
at most $\Ne$ elements with absolute values greater than $\alpha$.
Noticing
$\norm{\Ab(\sbb^0-\sbb^{\sigma_0})}{}=\norm{\Ab\sbb^0-\xb}{}<\epsilon$,
we see that $(\sbb^0-\sbb^{\sigma_0})$ satisfies the conditions of
Lemma~\ref{lem: noisy}, and hence:
\begin{eqnarray}
\norm{\sbb^0-\sbb^{\sigma_0}}{} \leq (M+1) (\Nu\alpha + \epsilon).
\end{eqnarray}
\end{proof}

{\bf Remark 1. \ } A few calculations show that the Gaussian
families (\ref{eq: Gaussian f}) satisfies the condition~1 of the
theorem for $\gamma=\exp(-1/2)$ and the condition~2 for
$\alpha=-\sigma_0\sqrt{2\ln(\nu)}$. Family of functions defined
by (\ref{eq: second order mobius f}) also satisfy the conditions
of this theorem.

\vspace{0.5em}

{\bf Remark 2. \ } Note that for Gaussian family of
functions and under the condition $k<\Ne/2$, accuracy of the
solution is proportional to the noise power\footnote{Optimal
choice of $\sigma_0$ is also proportional to the noise power.}. In
fact, we have accuracy of at least $C\cdot\epsilon$, where:
\begin{displaymath}
C=\big(\frac{\exp(-1/2)\Nu^2\sqrt{2\ln\Nu} \,
\norm{\Ab^T(\Ab\Ab^T)^{-1}}{}}{\Ne-2k}+1 \big)(M+1).
\end{displaymath}
If $\epsilon\rightarrow 0$, by choosing $\sigma_0$ according to
(\ref{eq: choose sigma noisy}), $\sbb^{\sigma_0}$ converges to
$\sbb^0$.

{\bf Remark 3. \ } According to Theorem~\ref{theo: noisy case}, in
contrast to the noiseless case, it is not possible here to achieve
arbitrarily accurate solutions. Accuracy is bounded by the noise
power, and to guaranty an error estimation less than $\beta$ using Theorem
\ref{theo: noisy case}, it is required to satisfy $\epsilon <
\beta / C$.

%%%%%%%%%%%%%%%%%%%%%%%%%%%%%%%%%%%%%%%%%%%%%%%%%%%%%%%%%%%%%%%%%%%
%%%%%%%%%%%                 New Section                 %%%%%%%%%%%
%%%%%%%%%%%%%%%%%%%%%%%%%%%%%%%%%%%%%%%%%%%%%%%%%%%%%%%%%%%%%%%%%%%
\section{Experimental Results} \label{sec: experiments}

In this section, the performance of the presented approach is
experimentally verified and is compared with BP (and with FOCUSS for
the first experiment). The effects of the
parameters, sparsity, noise, and dimension on the performance are
also experimentally discussed.

In all of the experiments (except in Experiment~3), sparse
sources are artificially created using a Bernoulli-Gaussian
model: each source is `active' with probability $p$, and is
`inactive' with probability $1-p$. If it is active, each sample is
a zero-mean Gaussian random variable with variance
$\sigma^2_\mathrm{on}$; if it is not active, each sample is
a zero-mean Gaussian random variable with variance
$\sigma^2_\mathrm{off}$, where $\sigma^2_\mathrm{off} \ll
\sigma^2_\mathrm{on}$. Consequently, each $s_i$ is distributed as:
\begin{equation}
\label{eq: the sources model}
s_{i}\sim p \cdot
\mathcal{N}(0,\sigma_{\mathrm{on}})+(1-p) \cdot
\mathcal{N}(0,\sigma_{\mathrm{off}}),
\end{equation}
where $p$ denotes the probability of activity of the sources, and
sparsity implies that $p \ll 1$. $\sigma_{\mathrm{off}}$ 
models the noise in the sources, that is, small values of the
sparse sources in their inactive case. This parameter is mostly
meaningful in SCA applications, in which, usually the sources in
their inactive states are not exactly zero. However, in sparse
decomposition applications $\sigma_{\mathrm{off}}$ can be usually
set to zero, that is, most elements of the dictionary are absent
in the decomposition.

In our simulations, $\sigma_{\mathrm{on}}$ is always fixed to 1.
The effect of $\sigma_{\mathrm{off}}$ is investigated only in the
first experiment. In all the other experiments it is set to
zero.

Each column of the mixing matrix is randomly generated using the
normal distribution and then is normalized to unity. Then, the
mixtures are generated using the noisy model:
\begin{equation}
\label{eq: noisy model}
\xb=\Ab\sbb+\nb,
\end{equation}
where $\nb$ is an additive white Gaussian noise (modeling sensor noise,
or decomposition inaccuracy) with covariance matrix $\sigma_n \Ib_\Ne$
(where $\Ib_\Ne$ stands for the $\Ne \times \Ne$ identity matrix).

To evaluate the estimation quality, Signal-to-Noise Ratio (SNR)
and Mean Square Error (MSE) are used. SNR (in dB) is defined as
$20 \log(\|\sbb\|/\|\sbb-\hsb\|)$ and MSE as
$(1/m)\|\sbb-\hsb\|^{2}$, where $\sbb$ and $\hsb$ denote the
actual source and its estimation, respectively.

Using (\ref{eq: the sources model}), the number of active sources
has a binomial distribution with average $\Nu p$. 
In the experiments, we will use the parameter $k=mp$, instead of $p$. 

\vspace{0.5em}

\noindent \textbf{Experiment 1. Performance analysis}\nopagebreak[4]

In this experiment, we study the computational cost of the
presented method, and compare its performance with
$\ell_1$-magic~\cite{CandR05} as one of the fastest
implementations of interior-point LP, and with FOCUSS\footnote{For FOCUSS,
we have used the
MATLAB code available at {\tt http://dsp.ucsd.edu/\~{}jfmurray/software.htm}}.
In rest of the paper, by LP we mean $\ell_1$-magic implementation of the
interior point LP.

\begin{table}[tb]
\centering%
\caption{%
    Progress of SL0 for a problem with %
    $\Nu = 1000$, $\Ne=400$ and $k = 100$ $(p = 0.1)$.%
}
\begin{tabular}{c | c | c | c }
\hline
itr. \# & $\sigma$ & MSE             &SNR (dB)  \\
\hline  1 & 1      & $4.84\,e\,{-2}$ &  2.82    \\
        2 & 0.5    & $2.02\,e\,{-2}$ &  5.19    \\
        3 & 0.2    & $4.96\,e\,{-3}$ & 11.59    \\
        4 & 0.1    & $2.30\,e\,{-3}$ & 16.44    \\
        5 & 0.05   & $5.83\,e\,{-4}$ & 20.69    \\
        6 & 0.02   & $1.17\,e\,{-4}$ & 28.62    \\
        7 & 0.01   & $5.53\,e\,{-5}$ & 30.85    \\
%        \multicolumn{5}{c}{}\\
        \hline \hline
        \multicolumn{1}{c}{algorithm} & \multicolumn{1}{c}{total
        time (sec)} & \multicolumn{1}{c}{MSE} & SNR (dB)\\
        \hline
        \multicolumn{1}{c}{SL0} & \multicolumn{1}{c}{$0.227$} &
        \multicolumn{1}{c}{$5.53\,e\,{-5}$} & 30.85\\
        \multicolumn{1}{c}{LP ($\ell_1$-magic)} & \multicolumn{1}{c}{$30.1$} &
        \multicolumn{1}{c}{$2.31\,e\,{-4}$} & 25.65 \\
        \multicolumn{1}{c}{FOCUSS} & \multicolumn{1}{c}{$20.6$} &
        \multicolumn{1}{c}{$6.45\,e\,{-4}$} & 20.93
\end{tabular}
\label{table: LPcomparison}
\end{table}

\begin{figure}[tb]
\begin{center}
\includegraphics[width=7cm]{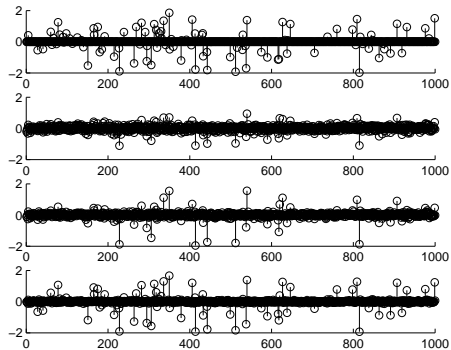}
\end{center}
\caption{Evolution of SL0 toward the solution:
$\Nu=1000$, $\Ne=400$ and $k=100 (p=0.1)$. From top to bottom, first
plot corresponds to the actual source, second plot is its estimation
at the first level ($\sigma=1$), third plot is its estimation at the
second level ($\sigma=0.5$), while the last plot is its estimation
at third level ($\sigma=0.2$).} \label{fig: LPcomparison}
\end{figure}

The values used for the first part of the experiment are
$\Nu=1000$, $\Ne=400$, $p=0.1$, $\sigma_{\mathrm{off}}=0$,
$\sigma_{\mathrm{on}}=1$, $\sigma_n=0.01$ and the sequence of
$\sigma$ is fixed to [1, 0.5, 0.2, 0.1, 0.05, 0.02, 0.01]. $\mu$
is fixed to $2.5$. For each value of $\sigma$ the
gradient-projection loop (the internal loop) is repeated three
times, \ie $L=3$ (influence of L is discussed in part of
experiment 2; in all other experiments $\mu$ and $L$ are
fixed to 2.5 and 3).

We use the CPU time as a measure of complexity. Although it 
is not an exact measure, it gives a rough estimation of
the complexity, for comparing SL0 and LP algorithms. Our
simulations are performed in MATLAB7 environment using an AMD
Athlon sempron 2400+, 1.67GHz processor with 512MB of memory, and
under Microsoft Windows XP operating system.

Table~\ref{table: LPcomparison} shows the gradual improvement in
the output SNR after each iteration, for a typical run of SL0.
Moreover, for this run, the total time and final SNR have been
shown for SL0, for LP, and for FOCUSS. It is seen that SL0
performs {\em two orders of magnitude faster than LP, while it
produces a better SNR} (in some applications,
it can be even {\em three orders of magnitudes}
faster: see Experiment~6). Figure~\ref{fig: LPcomparison} shows the
actual source and it's estimations at different iterations for
this run of SL0.

The experiment was then repeated 100 times (with the same
parameters, but for different randomly generated sources and
mixing matrices) and the values of SNR (in dB) obtained over these
simulations were averaged. These averaged SNR's for SL0, LP, and FOCUSS
were respectively 30.85dB, 26.70dB, and 20.44dB; with respective
standard deviations 2.36dB, 1.74dB and 5.69dB. The minimum values of
SNR for these methods were respectively 16.30dB, 18.37dB, and 10.82dB.
Among the 100 runs of the algorithm, the number
of experiments for
which SNR$>$20dB was 99 for SL0 and LP, but only 49 for FOCUSS.

In the second part of the experiment, we use the same parameters as in the first part, except $\sigma_{\mathrm{off}} = 0.01$ to model the noise of the
sources in addition to AWG noise modeled by $\sigma_n$.
The averaged SNR's for SL0, LP, and FOCUSS
were respectively 25.93dB, 22.15dB and 18.24; with respective
standard deviations 1.19dB, 1.23dB and 3.94dB.

\vspace{0.5em}

\noindent \textbf{Experiment 2. Dependence on the parameters}\nopagebreak[4]

In this experiment, we study the dependence of the
performance of SL0 to its parameters. The sequence of $\sigma$ is
always chosen as a decreasing geometrical sequence
$\sigma_j=c \sigma_{j-1}, \, 1\leq j \leq J$, which is determined by
the first and last elements, $\sigma_1$ and $\sigJ$, and the scale factor $c$.
Therefore, when considering the effect of the sequence of
$\sigma$, it suffices to discuss the effect of these three parameters
on the performance. Reasonable choice of $\sigma_1$, and also
approximate choice of $\mu$ have already been discussed in
Remarks 2 to 5 of Section~\ref{sec: alg}. Consequently, we are
mainly considering the effects of other parameters.

The general model of the sources and the mixing system, given by
(\ref{eq: the sources model}) and (\ref{eq: noisy model}), has
four essential parameters: $\sigma_{\mathrm{on}}$, 
$\sigma_{\mathrm{off}}$%
, $\sigma_n$%
, and $p$%
. We can
control the degree of source sparsity and the power of the noise
by changing\footnote{Note that the sources are generated using
the model (\ref{eq: the sources model}). Therefore, for example
$k=100$ does not necessarily mean that exactly 100 sources are
active.} $k=\Nu p$ and $\sigma_n$. We examine the performance of
SL0 and its dependence to these parameters for different levels of
noise and sparsity. In this and in the followings,  except
Experiment 6, all the simulations are repeated 100 times
with different randomly generated sources and mixing matrices and
the values of the SNR's (in dB) obtained over these simulations
are averaged.

\begin{figure}
\begin{center}
   \renewcommand{\subfigcapskip}{-5.8cm}
   \renewcommand{\subfigtopskip}{5pt}
   \renewcommand{\thesubfigure}{{\hspace*{3em}(\alph{subfigure})}}
   \subfigure[]{
   %\label{fig: cfuncnoise}
   \includegraphics[height=5.5cm]{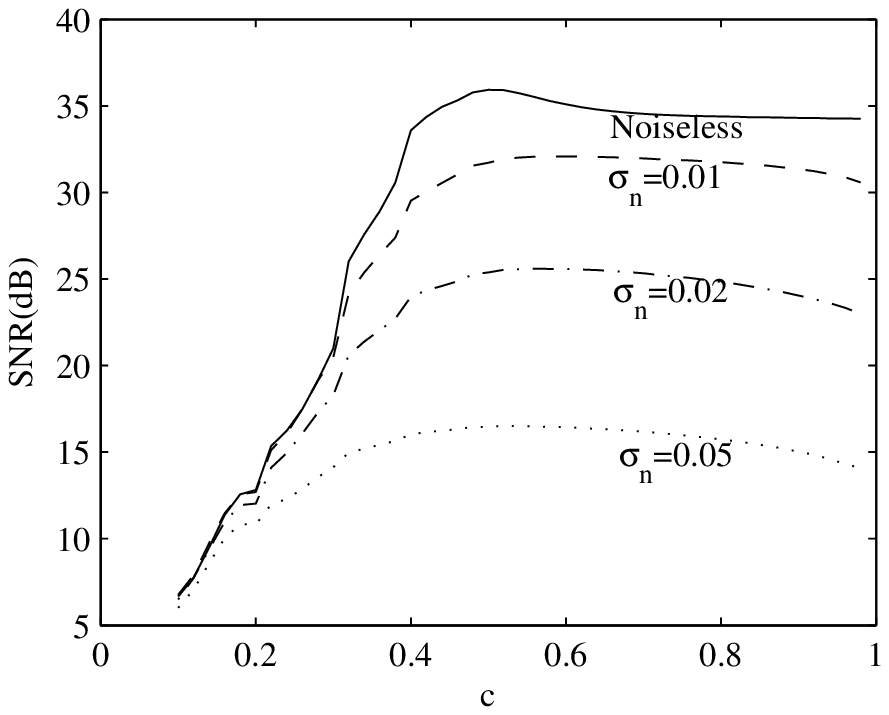}}
   \hskip 0.01\textwidth
   \subfigure[]{
   %\label{fig: cfuncsparsity}
   \includegraphics[height=5.5cm]{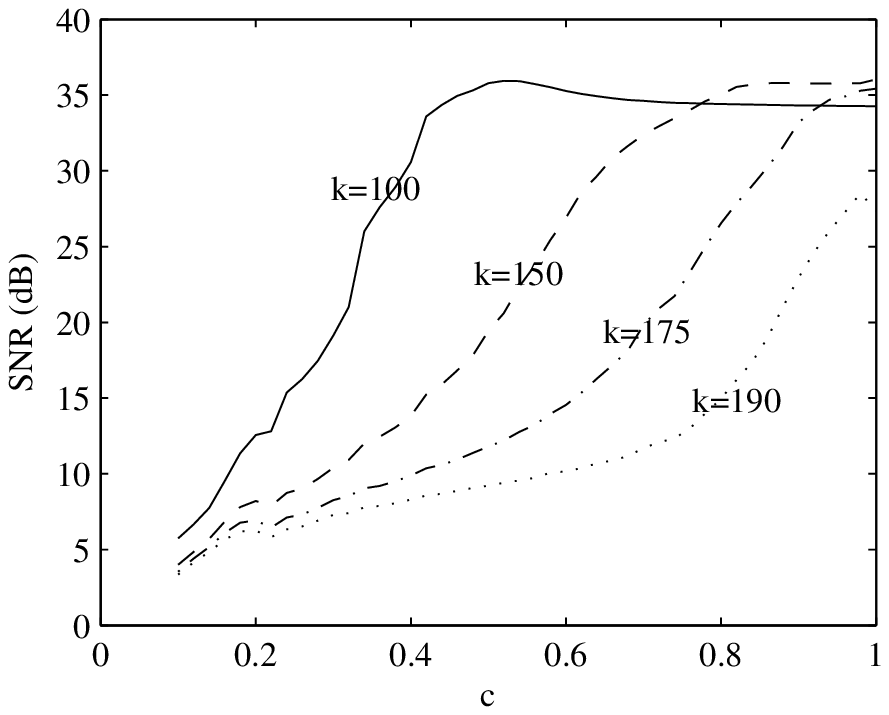}}
\end{center}

\caption{Performance of SL0 as function of $c$ for
the case $\Nu=1000$ and $\Ne=400$ (SNR's are averaged over 100 runs
of the algorithm). $\sigma_1$ is fixed to 1 (large
enough) and $\sigJ$ is fixed to 0.01 (small enough). In (a) $k$
is fixed to 100 and effect of noise is investigated. In (b)
$\sigma_n$ is fixed to 0.01 and effect of sparsity factor is
analyzed.}
 \label{fig: SNR versus Noise and Sparsity}
\end{figure}

Figures \ref{fig: SNR versus Noise and Sparsity} represents the
averaged SNR (as the measure of performance) versus the scale
factor $c$, for different values of $k=\Nu p$ and $\sigma_n$. It
is clear from Fig. \ref{fig: SNR versus Noise and Sparsity}(a) that SNR increases when
$c$ increases form zero to one. However, when $c$ exceeds a
critical value (0.5 in this case), SNR remains constant and does
not increase anymore.

Generally, the optimal choice of $c$ depends on the application.
When SNR is the essential criterion, $c$ should be chosen large,
resulting in a more slowly decreasing sequence of $\sigma$, 
and hence in a higher
computational cost. Therefore, the choice of $c$ is a trade-off
between SNR and computational cost. However, as seen in the
figures, when $c$ approaches to unity, SNR does not increase infinitely.
In Fig.~\ref{fig: SNR versus Noise and Sparsity}(a),  the optimal value of 
$c$, i.e. the smallest value of $c$ that achieves the maximum SNR, is 
approximatively $c=0.5$.  However, it is clear from 
Fig.~\ref{fig: SNR versus Noise and Sparsity}(b)
that the optimal choice of $c$ depends on the sparsity,
but not on the noise power. Exact calculation of the optimal $c$
might be very hard. To guarantee an acceptable performance, it
suffices to choose $c$ greater than its optimal value.

From \cite{Dono04}, we know that $k<\Ne/2$ is a theoretical limit
for sparse decomposition. However, most of the current methods cannot approach
this limit (see Experiment 3). In Fig. \ref{fig: SNR versus Noise and Sparsity}(b),
$k=190\simeq200=\Ne/2$ is plotted, and it is clear that by
choosing $c$ larger than $0.9$ an acceptable performance can be
achieved (however, with a much higher computational cost).

\begin{figure}
\begin{center}
   \renewcommand{\subfigcapskip}{-5.8cm}
   \renewcommand{\subfigtopskip}{5pt}
   \renewcommand{\thesubfigure}{{\hspace*{3em}(\alph{subfigure})}}
   \subfigure[]{
   %\label{fig: sigendnoise}
   \includegraphics[height=5.5cm]{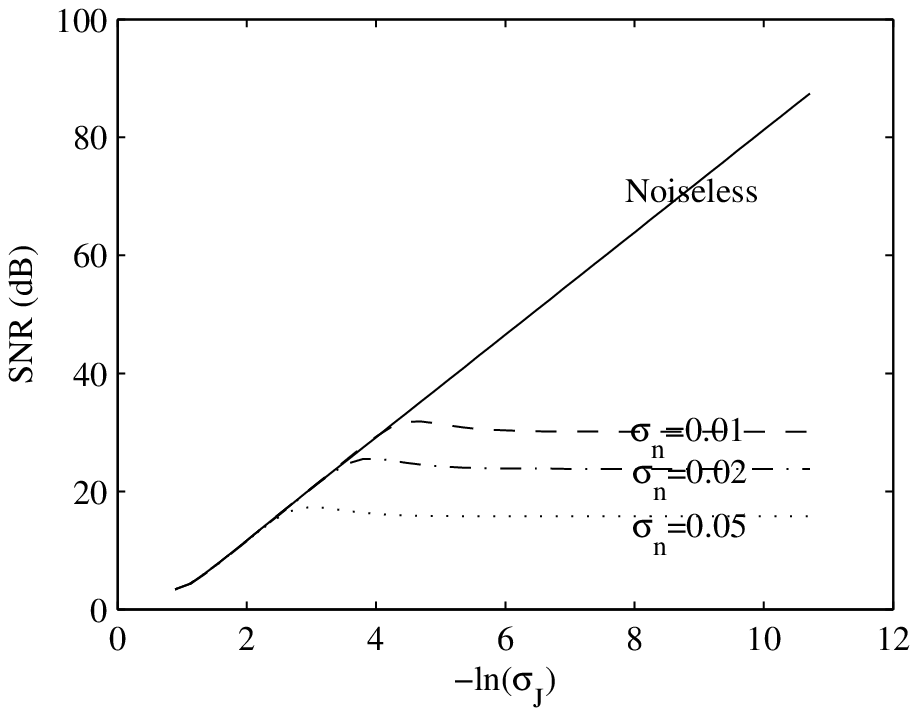}}
   \hskip 0.01\textwidth
   \subfigure[]{
   %\label{fig: sigendsparsity}
   \includegraphics[height=5.5cm]{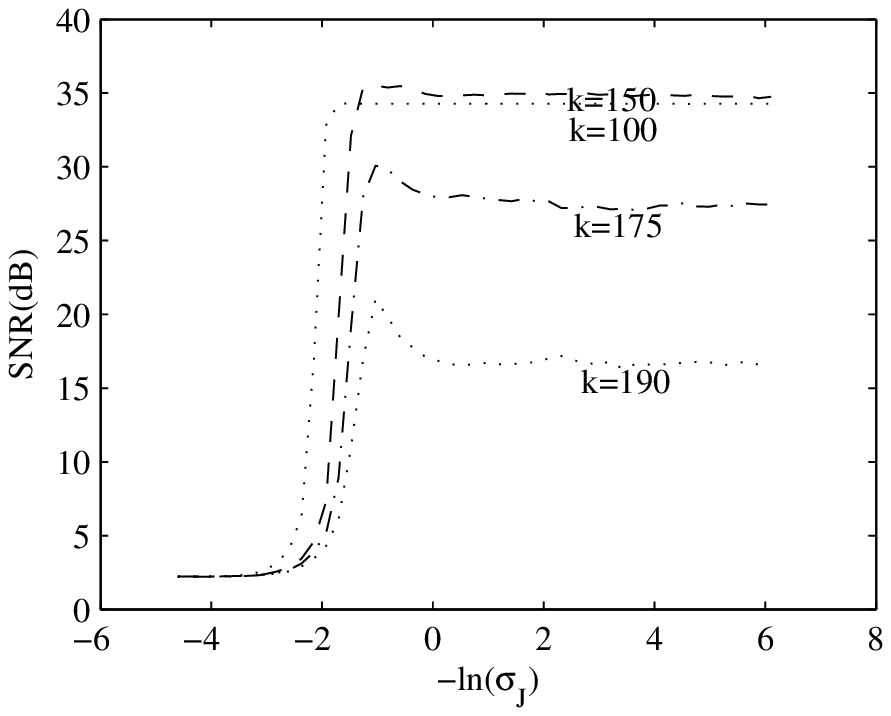}}
\end{center}

\caption{Performance of SL0 versus $\sigJ$
for $\Nu=1000$ and $\Ne=400$ (SNR's are averaged over 100 runs
of the algorithm). $\sigma_1$ is fixed to 1
(large enough) and $c$ is fixed to 0.8 (near enough to one). In (a)
$k$ is fixed to 100 and effect of noise is investigated. In (b)
$\sigma_n$ is fixed to 0.01 and effect of sparsity factor is
analyzed.}
\label{fig: SNR vs sigma and sparsity}
\end{figure}

In Fig. \ref{fig: SNR vs sigma and sparsity},
SNR is plotted versus $-\ln(\sigJ)$ (where $\sigJ$
is the last and smallest $\sigma$) for
different values of $k$ and $\sigma_n$. 
In Fig. \ref{fig: SNR vs sigma and sparsity}(a), for the noiseless case, SNR
increases linearly, by increasing in $-\ln(\sigJ)$. Although
not directly clear from the figure, calculation of the obtained
values of the figure better shows this linear relationship. This
confirms the results of Theorem \ref{theo: solution convergence}
(accuracy is proportional to the final value of $\sigma$). In the
noisy case, SNR increases first, and then remains constant. As was
predicted by Theorem \ref{theo: noisy case}, in the noisy case the
accuracy is bounded and might not be increased arbitrarily.

Generally, the optimal choice of $\sigJ$ depends on the
application. In applications in which SNR is highly more
important than the computational load, $\sigJ$ should be chosen
small, resulting in a larger sequence of $\sigma$, and hence a higher 
computational cost. However, excessively small choice of $\sigJ$ (smaller
than the optimal choice) does not improve SNR (in fact SNR is
slightly decreased. Recall also the Remark~6 of Section~\ref{sec:
alg}). 
It is clear from Fig.~\ref{fig: SNR vs sigma and sparsity} that
the optimal choice of $\sigJ$ depends on the noise power, but
not on the sparsity. Exact calculation of the optimal $\sigJ$
might be very hard. To guarantee an acceptable performance, it
suffices to choose $\sigJ$ less than its optimal value.

From this experiment it can be concluded that, although finding
optimal values of the parameters for optimizing the SNR with the least
possible computational cost may be very hard, the algorithm is
not very sensitive to the parameters, and it is not difficult to choose
a sequence of $\sigma$ (\ie $c$ and $\sigJ$).

\begin{figure}
\centering

 \includegraphics[width=8cm]{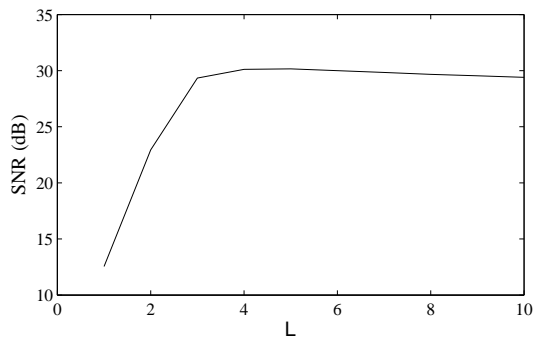}

\caption{Averaged SNR (on 100 runs of the algorithm) versus $L$
for the case $\Nu=1000$ and
$\Ne=400$, $k=100$ and $\sigma_n=0.01$}

 \label{fig: SNR and Time versus L}
\end{figure}

Finally, to study the effect of $L$ (number of iterations of the
internal steepest ascent loop), the parameters are fixed to the 
values used at the beginning of Experiment~1, and the averaged SNR
(over 100 runs of the algorithm) is plotted versus $L$
in Fig. \ref{fig: SNR and Time versus L}. 
It is clear from this figure that the final SNR
achieves its maximum for a small $L$, and no longer improves
by increasing it, while the computation cost is directly
proportional to $L$. Hence, as it was said in Remark~1 of 
Section~\ref{sec: alg} and Remark~3 of
Section~\ref{sec: conv to sol}, we generally fix $L$ to a small
value, say $L=3$.

\vspace{0.5em}

\noindent \textbf{Experiment 3. Effect of sparsity on the performance}\nopagebreak[4]

How much sparse a source vector $\sbb$ should be to make its estimation
possible using our algorithm? Here, we try to
answer this question experimentally. As mentioned before, there is
the theoretical limit of $\Ne/2$ on the maximum number of active
sources to insure the uniqueness of the sparsest solution. But,
practically, most algorithms cannot achieve this limit~\cite{Dono04,GoroR97}.

\begin{figure}
\begin{center}
\includegraphics[width=7cm]{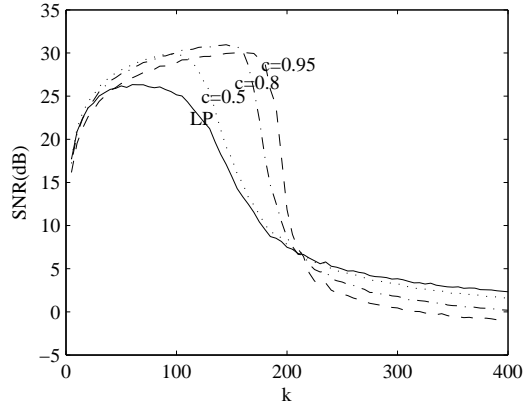}
\end{center}
\caption{Averaged SNR's (over 100 runs of the algorithm)
versus $k$, the average number
of active sources, for SL0 algorithm with several values of $c$, and for
LP. The parameters are $\Nu=1000$, $\Ne=400$, $\sigma_1=1$, $\sigJ=0.01$, 
$\sigma_n=0.01$.} 
\label{fig: sep}
\end{figure}

To be able to measure the effect of sparsity, instead of
generating the sources according to the model (\ref{eq: the
sources model}), we randomly activate exactly $k$ elements out of $\Nu$
elements. Figure~\ref{fig: sep} then shows the output SNR 
versus $k$, for
several values of $c$, and compares the results with LP. Note
that SL0 outperforms LP, specially in cases where
$k\simeq\Ne/2=200$.

It is obvious from the figure that all methods work well if
$k$ is smaller than a critical value, and they start breaking
down as soon as $k$ exceeds this critical value. Figure~\ref{fig: sep} shows
that the break-down value of $k$ for LP and for SL0 with $c=0.5$
is approximately 100 (half of the theoretical limit $\Ne/2=200$).
For $c=0.8$ and $c=0.95$, this break-down value is
approximately 150 and 180. Consequently, with our algorithm, it
is possible to estimate less sparse sources than with LP algorithm. 
It seems also that by
pushing $c$ toward 1, we can push the breaking-down point toward
the theoretical limit $n/2$; however, the
computational cost might become intolerable, too.

\vspace{0.5em}

\noindent \textbf{Experiment 4. Robustness against noise}\nopagebreak[4]

\begin{figure}
\begin{center}
\includegraphics[width=7cm]{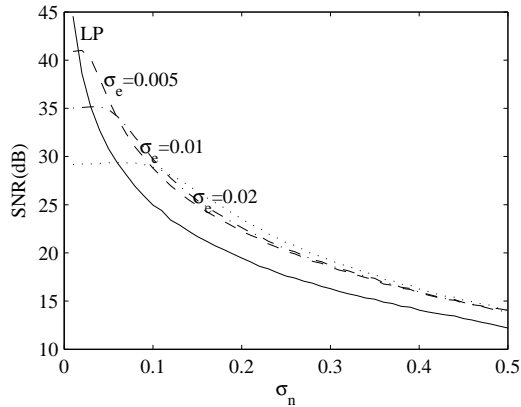}
\end{center}
\caption{Averaged SNR's (over 100 runs of the algorithm)
versus the noise power
$\sigma_n$ for different values of $\sigJ$, and for LP. The
parameters are $\Nu=100$, $\Ne=400$, $k=100$, $\sigma_1=1$, and
$c=0.8$.} \label{fig: noisefuncsigend}
\end{figure}

In this experiment, the effect of the noise variance,
$\sigma_n$, on the performance is investigated for different
values of $\sigJ$ and is compared with the performance of LP.
Figure~\ref{fig: noisefuncsigend} depicts SNR versus
$\sigma_n$ for different values of $\sigJ$ for both methods.
The figure shows the robustness of SL0 against small values of
noise. In the noiseless case ($\sigma_n<.02$), LP performs better
(note that $\sigma_\mathrm{off}=0$, and in SL0, $\sigma$ is decreased only to 0.005).
In the noisy case, smoothed-$\ell^0$ achieves better SNR. Note
that the dependence of the optimal $\sigJ$ to $\sigma_n$ is
again confirmed by this experiment.

\vspace{0.5em}
\noindent \textbf{Experiment 5. Number of sources and sensors}\nopagebreak[4]

In this experiment, we investigate the effect of the
system scale (\ie the dimension of the mixing matrix, $\Nu$ and
$\Ne$) on the performance and justify the scalability of SL0.

\begin{figure}
\begin{center}
   \renewcommand{\subfigcapskip}{-5.8cm}
   \renewcommand{\subfigtopskip}{5pt}
   \renewcommand{\thesubfigure}{{\hspace*{3em}(\alph{subfigure})}}
   \subfigure[]{
   \includegraphics[height=5.5cm]{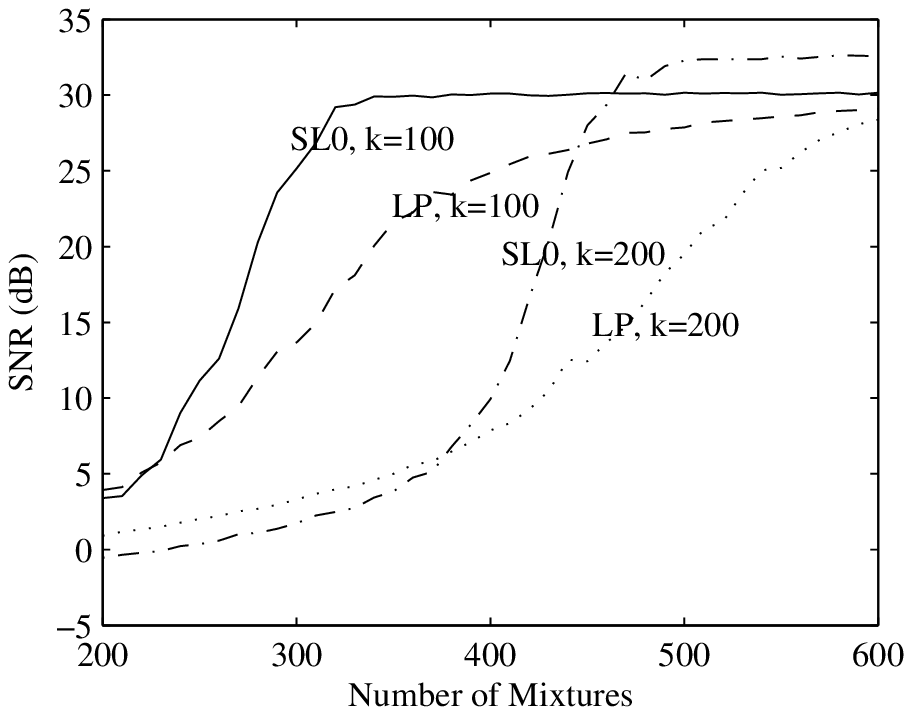}}
   \hskip 0.01\textwidth
   \subfigure[]{
   \includegraphics[height=5.5cm]{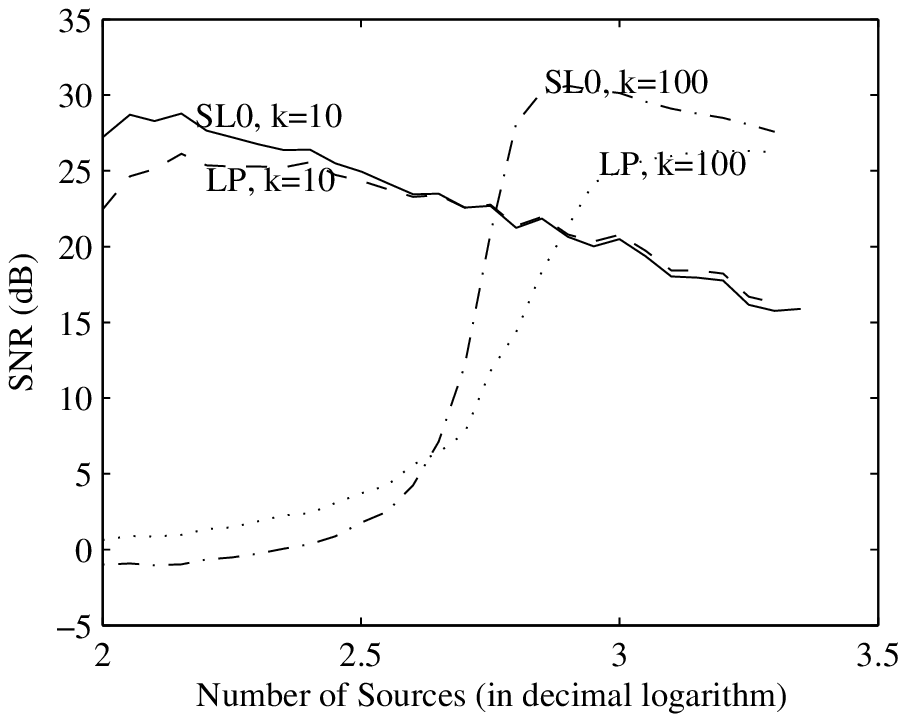}}
\end{center}

\caption{Effect of scale on performance (SNR's are averaged over 100 runs
of the algorithm). $\sigma_n=0.01$, $c = 0.8$, $\sigma_1=1$,
$\sigJ=0.01$, and SL0 is compared with LP. In (a) $\Nu$ is
fixed to 1000 and SNR is plotted versus $\Ne$ for
different values of $k$. In (b)
SNR is plotted versus $\log(\Nu)$ for different values of
$k$, while $\Ne$ is fixed to $\lceil 0.4 \Nu \rceil$.}
\label{fig: scale effect}
\end{figure}

First, to analyze the effect of the number of mixtures ($\Ne$), by fixing
$\Nu$ to 1000, SNR is plotted versus $\Ne$, for
different values of $k$ in Fig.~\ref{fig: scale effect}(a). It is clear
from this figure that both methods perform poorly while $2k>\Ne$
(note that the sparsest solution is not necessarily unique in
this case). SL0 performs better as soon as $\Ne$ exceeds $2k$ (the
theoretical limit for the uniqueness of the sparsest solution).

Then, to analyze the effect of scale, $\Ne$ is fixed to $\lceil 0.4
\Nu \rceil$, and SNR is plotted versus $\log(\Nu)$ for
different values of $k$ in Fig. \ref{fig: scale effect}(b). From this figure it
is obvious that SL0 and LP perform similarly for small values of $k$
($k\simeq10)$, but SL0 outperforms LP for larger values of $k$ ($k\simeq100$).

\vspace{0.5em}
\noindent \textbf{Experiment 6. Computational Cost in BSS applications}\nopagebreak[4]

In BSS and SCA applications, the model (\ref{eq: noisy
model}) is written as $\xb(t)=\Ab \sbb(t) + \nb(t), \, 1 \le t \le
T$, where $T$ is the number of samples. In matrix form, this can
be written as $\Xb = \Ab \Sb + \Nb$, where $\Xb$, $\Sb$, and $\Nb$
are respectively $\Ne \times T$, $\Nu \times T$ and $\Ne \times
T$ matrices, where each column stands for a time sample.

For solving this problem with LP, the system $\xb(t)=\Ab \sbb(t)
+ \nb(t)$ should be individually solved for each value of $1\le
t\le T$. This trivial approach can also be used with SL0. However,
since all the steps of SL0 presented in Fig.~\ref{fig: the alg} are
in matrix form, it can also be directly run on the whole matrices
$\Xb$ and $\Sb$. Because of the speed of the current matrix
multiplication algorithms\footnote{Let $\Ab$, $\sbb$ and $\Sb$ be
$\Ne \times T$, $\Nu \times 1$ and $\Nu \times T$ matrices,
respectively. In MATLAB, the time required for the multiplication
$\Ab \Sb$ is highly less than $T$ times of the time required for
the multiplication $\Ab \sbb$. This seems to not be due to the MATLAB's 
interpreter, but a property of Basic Linear Algebra
Sub-programs (BLAS). BLAS is a free set of highly optimized
routines for matrix multiplications, and is used by MATLAB for its
basic operations. This property does not exist in MATLAB~5.3
which was not based on BLAS.}, this results in an increased speed in the
total decomposition process.

\begin{figure}
\begin{center}
\includegraphics[width=7cm]{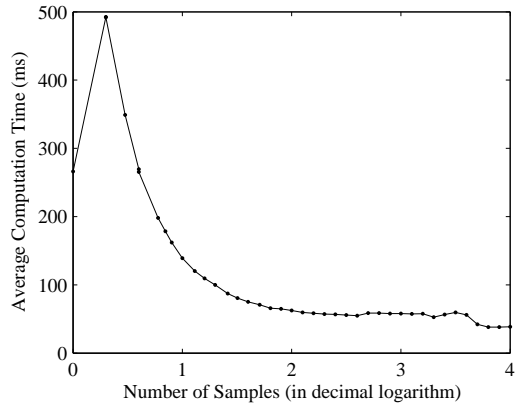}
\end{center}
\caption{Average computation time per sample of SL0, as a
function of $T$, number of (time) samples, for the case $\Nu=1000$, $\Ne=400$ and
$k=100$. $\sigma_n$ is chosen 0.01 and the sequence of $\sigma$ is fixed to
[1, 0.5, 0.2, 0.1, 0.05, 0.02, 0.01], the same parameter used in first experiment.} \label{fig: BSS}
\end{figure}

Figure \ref{fig: BSS} shows the average computation time per
sample of SL0 for a single run of the algorithm, as a function of
$T$ for the case $\Nu=1000$, $\Ne=400$ and $k=100$. The figure
shows that by increasing $T$, average computation time first
increases, then decreases and reach to a constant. For $T=1$, the
computation time is 266ms (this is slightly different with the time of
the first experiment, 227ms, because these are two different runs).
However, for $T=10000$, the average computation time per sample decreases to 38ms.
In other words, in average, SL0 finds the sparse solution of a linear system
of 400 equations and 1000 unknowns just in 38ms (compare this with
30s for $\ell_1$-magic, given in Experiment 1).

%%%%%%%%%%%%%%%%%%%%%%%%%%%%%%%%%%%%%%%%%%%%%%%%%%%%%%%%%%%%%%%%%%%
%%%%%%%%%%%                 New Section                 %%%%%%%%%%%
%%%%%%%%%%%%%%%%%%%%%%%%%%%%%%%%%%%%%%%%%%%%%%%%%%%%%%%%%%%%%%%%%%%
\section{Conclusions}
\label{sec: con} In this paper, we showed that the smoothed \Lzero\ norm
can be used for finding sparse solutions of an USLE. We showed also that the smoothed
version of the \Lzero\ norm not only solves the problem of intractable computational
load of the minimal \Lzero\ search, but also results in an algorithm
which is highly faster than the state-of-the-art algorithms based
on minimizing the \Lone\ norm. Moreover, this smoothing solves the problem
of high sensitivity of \Lzero\ norm to noise. In another point of view, the smoothed
\Lzero\ provides a smooth measure of sparsity.

The basic idea of the paper was justified by both theoretical and
experimental analysis of the algorithm. In the theoretical part,
Theorem \ref{theo: solution convergence} shows that SL0 is
equivalent to $\ell^0$-norm for a large family of functions
$f_\sigma$. Theorem \ref{theo: min L2 norm} gives a strong
assessment for using $\ell^2$-norm solution for initialization.
This theorem also suggests that the minimal \Ltwo\ norm can
be seen as a rough estimation of the sparse solution (like Method Of
Frames), which will be modified in the future iterations.
Theorem \ref{theo: noisy case} justifies the robustness of SL0
against noise.

Other properties of the algorithm were studied experimentally. In
particular, we showed that (1) the algorithm is highly faster than the
state-of-the-art LP approaches (and it is even more efficient
in SCA applications), (2) choosing suitable
values for its parameters is not difficult, (3) contrary to
previously known approaches it can work if the
number of non-zero components of $\sbb$ is near
$n/2$ (the theoretical limit for the uniqueness of
the sparse solution), and (4) the algorithm is robust
against noise.

Up to now, we have no theoretical result for determining how 
much `gradual' we should decrease the sequence of
$\sigma$, and it remains an open problem for future works. 
Some open questions related to this issue are: Is there any sequence of $\sigma$ which
guaranties escaping from local maxima
for the Gaussian family of functions $F_\sigma$ 
given in~(\ref{eq: Gaussian f})?  If yes, how to find this sequence? 
If not, what happens with other families of functions $F_\sigma$? Moreover, 
is there any (counter-)example of $\Ab$, $\sbb$ and $\xb$ for which we can prove that for {\em any}
sequence $\sigma$ the algorithm will get trapped into a local maximum?
These issues, mathematically difficult but essential for proving algorithm convergence, 
are currently investigated. 
However, Experiment 2 showed that it is fairly easy
to set some parameters to achieve a suitable performance. 
Moreover, for an estimation $\hsb$ of the 
sparsest source (obtained by any method), we provided in Remark~5 
of Section~\ref{sec: conv to sol} an upper bound for the estimation error. 

In addition, future works include better treatment of the noise
in the model (\ref{eq: noisy model}) by taking it directly into
account in the algorithm ({e.g.} by adding a penalty term to
$F_\sigma$). Moreover, testing the algorithm on
different applications (such as compressed sensing) using
real-world data is under study in our group.

%%%%%%%%%%%%%%%%%%%%%%%%%%%%%%%%%%%%%%%%%%%%%%%%%%%%%%
%\bibliography{SepSrc}
%\bibliographystyle{IEEEbib}

%%%%%%%%%%%%%%%%%%%%%%%%%%%%%%%%%%%%%%%%%%%%%%%%%%%%%%

\end{document}